\documentclass[preprint]{aastex}
\usepackage{gensymb} 
\usepackage{ amssymb }

\newcommand{\fw}{FWHM}

\newcommand{\ch}{CH$_{3}$OH}
\newcommand{\nhh}{NH$_{3}$}

\newcommand{\water}{H$_{2}$O}

\def\lesssim{\mathrel{\hbox{\rlap{\hbox{\lower4pt\hbox{$\sim$}}}\hbox{$<$}}}}
\def\gtsim{\mathrel{\hbox{\rlap{\hbox{\lower4pt\hbox{$\sim$}}}\hbox{$>$}}}}

\def\nh3{\mbox{${\rm NH_3}$}}

\slugcomment{July 14th, 2010}

\shorttitle{The Spitzer ice legacy}
\shortauthors{\"Oberg et al.}

\begin{document}

\title{The Spitzer ice legacy: Ice evolution from cores to protostars}

\author{
Karin I. \"Oberg\altaffilmark{1,2}}
\affil{Harvard-Smithsonian Center for Astrophysics, 60 Garden St, Cambridge, MA 02139, USA} 

\author{A.~C. Adwin Boogert}
\affil{IPAC, NASA Herschel Science Center, California Institute of Technology, Pasadena, CA 91125, USA} 

\author{Klaus M. Pontoppidan}
\affil{Space Telescope Science Institute, 3700 San Martin Drive, Baltimore, MD 21218, USA}

\author{ Saskia van den Broek,
Ewine F. van Dishoeck\altaffilmark{3}}
\affil{Leiden Observatory, Leiden University, PO Box 9513, 2300 RA Leiden, the Netherlands}

\author{Sandrine Bottinelli\altaffilmark{2}}
\affil{Centre d'Etude Spatiale des Rayonnements (CESR), CNRS-UMR 5187, 9 avenue du Colonel Roche, BP 4346, 31028 Toulouse Cedex 4, France}

\author{
Geoffrey A. Blake}
\affil{California Institute of Technology, Division of Geological and Planetary Sciences, Pasadena, CA 91125, USA}

\author{Neal J. Evans II}
\affil{Department of Astronomy, University of Texas at Austin, 1 University Station C1400, Austin, TX 78712, USA}

\altaffiltext{1}{Hubble Fellow}
\altaffiltext{2}{Leiden Observatory, Leiden University, PO Box 9513, 2300 RA Leiden, the Netherlands}
\altaffiltext{3}{Max Planck Institut f\"ur Extraterrestrische Physik (MPE), Giessenbachstr.1, 85748 Garching, Germany}

\begin{abstract}
\noindent Ices regulate much of the chemistry during star formation and account for up to 80\% of the available oxygen and carbon. In this paper, we use the {\it Spitzer} $c2d$ ice survey, complimented with data sets on ices in cloud cores and high-mass protostars, to determine standard ice abundances and to present a coherent picture of the evolution of ices during low- and high-mass star formation. The median ice composition H$_2$O:CO:CO$_2$:CH$_3$OH:NH$_3$:CH$_4$:XCN is 100:29:29:3:5:5:0.3 and 100:13:13:4:5:2:0.6 toward low- and high-mass protostars, respectively, and 100:31:38:4:--:--:-- in cloud cores.  In the low-mass sample, the ice  abundances with respect to H$_2$O of CH$_4$, NH$_3$, and the component of CO$_2$ mixed with H$_2$O  typically vary by $<$25\%, indicative of co-formation with H$_2$O. In contrast, some CO and CO$_2$ ice components, XCN and CH$_3$OH vary by factors 2--10 between the lower and upper quartile.  The XCN band correlates with CO, consistent with its OCN$^{-}$ identification. The origin(s) of the different levels of ice abundance variations are constrained by comparing ice inventories toward different types of protostars and background stars, through ice mapping, analysis of cloud-to-cloud variations, and ice (anti-)correlations. Based on the analysis, the first ice formation phase is driven by hydrogenation of atoms, which results in a H$_2$O-dominated ice. At later prestellar times, CO freezes out and variations in CO freeze-out levels and the subsequent CO-based chemistry can explain most of the observed ice abundance variations. The last important ice evolution stage is thermal and UV processing around protostars, resulting in CO desorption, ice segregation and formation of complex organic molecules. The distribution of cometary ice abundances are consistent with with the idea that most cometary ices have a protostellar origin.
\end{abstract}


\section{Introduction}

\noindent Star formation begins with the collapse of an interstellar cloud core to form a protostar, which grows from infall of envelope material and later through an accretion disk, where planets may  form. This physical evolution from cores to planetary systems is accompanied by a chemical evolution, which will affect planet and planetesimal compositions. Much of this chemical evolution takes place in icy mantles on interstellar grain surfaces, and the aim of this paper is to consolidate the knowledge of ice evolution during star formation coming out of the {\it Spitzer} Space Telescope mission.

The first ices were detected in the interstellar medium almost 40
years ago \citep{Gillet73} and from previous work, mostly toward
high-mass protostars, H$_2$O, CO and CO$_2$ ices are known to be
common during the cold and dense stages of star formation, with
abundances reaching 10$^{-4}$ $n_{\rm H_2}$. The ice mantles also
contain smaller amounts of CH$_3$OH, CH$_4$, NH$_3$ and XCN
\citep[e.g][]{Gibb04}. CH$_3$OH and other
ices are proposed sources of complex organic molecules
\citep{Charnley92,Garrod08,Oberg09d} and determining ice abundances
and production channels in star forming regions is therefore of great 
interest for studies of prebiotic chemistry. Most ices except for CO are predicted to form {\it in situ}
on interstellar grain surfaces through hydrogenation and oxygenation
of atoms and small molecules \citep{Tielens82}. Heat and UV radiation from protostars may result in
additional ice processing 
and \citet{Gibb04} suggested that some
CO$_2$ ice, and all CH$_3$OH and XCN ice form through such
processing based on observations toward high-mass protostars with the {\it Infrared Space Observatory} (ISO). This has been challenged by
observations of abundant CH$_3$OH and XCN ice toward low-mass
protostars \citep{Pontoppidan03,vanBroekhuizen05}, where ices are
protected from strong UV fields during most of their life time.  

\textit{Spitzer}'s high sensitivity made it possible to observe ices
toward low mass protostars and background stars at 5--30~$\mu$m.
The $c2d$ program \citep{Evans03} obtained IRS spectra
of 50 low-mass protostars 
\citep[][from now on Paper I-IV]{Boogert08, Pontoppidan08, Oberg08,
 Bottinelli10}, providing an unprecedented sample size that
complements the {\it ISO} data on
high-mass protostars. Additional {\it Spitzer} ice observations exist on ices toward
background sources, looking through molecular clouds at a range of
extinctions \citep{Bergin05,Knez05,Pontoppidan06,Whittet09,Boogert11} and toward
protostars in a high-UV environment
\citep{Reach09}. 

Because the  {\it Spitzer} spectrometer had a lower cutoff at $\lambda \sim 5$~$\mu$m, complete ice inventories can only be obtained by adding complementary ground-based spectroscopy to 
cover the strong 3 $\mu$m H$_2$O, the 4.65
$\mu$m CO, the 4.6 $\mu$m XCN, and the 3.53 $\mu$m CH$_3$OH features. The latter feature has been used to validate the use of the 9.7 $\mu$m feature to derive CH$_3$OH ice abundances (Paper I). The XCN feature has been shown empirically to consist of two components, whose relative contributions to the feature vary from source to source \citep{vanBroekhuizen05}. One of the components can be securely assigned to OCN$^-$ from comparison with laboratory spectra, while the carrier of the second component, peaking at 2175 cm$^{-1}$, is unknown. Suggested carriers of the 2175 cm$^{-1}$ component include OCN$^-$ present in a different ice environment, resulting in a different spectrum, as well as completely different molecules \citep{Pendleton99,vanBroekhuizen04}.

In the analysis of astrophysical ice spectra, comparison with laboratory ice spectra is key, both to identify ice species (such as OCN$^-$ above) and to characterize the ice morphology; i.e., some ice bands, such as CO and CO$_2$, cannot be fitted by a single laboratory ice mixture, but seem to trace molecules in two or more ice phases \citep[][Paper II]{Tielens91, Chiar95,Dartois99}. Traditionally the observed spectra have been directly compared to a superposition of pure and mixed ice spectra, characterizing the ice morphology in each source independently \citep[e.g.][]{Merrill76,Gibb04, Zasowski09}. The constraints are often degenerate, however, since ice spectral features vary with ice composition, temperature and radiation processing. To address this, ice bands have instead been decomposed phenomenologically into a small set of unique spectral components \citep[e.g.][ and \S2.2]{Tielens91, Pendleton99, Keane01,Pontoppidan03b}. The contributions of the derived components are then used to characterize the spectral bands in all sources in the sample. This method was employed in papers I and II. Because all observed spectra are decomposed into the same, small number of components this method provides information on the sample as a whole, i.e. it directly shows which parts of the spectral profile are ubiquitous and which are environment dependent. This is crucial information when assigning a component carrier -- without this, the degeneracy is almost always too large to draw conclusions about the structure of the ice from spectral profiles alone. 

Building on the analysis in Paper I-IV, this paper aims to establish an ice evolution scenario that also takes into account the present knowledge of ice abundances toward cloud cores \citep{Boogert11} and high-mass protostars \citep{Gibb04}. Section \ref{sec:obs} summarizes the sample characteristics. Section \ref{sec:res} presents new ice data for 10 low-mass protostars, followed by ice abundance medians toward low- and high mass protostars. The variation among the low-mass protostellar abundances are investigated followed by comparison of this ice sample with ices toward high-mass protostars, low-mass protostars in a high-radiation environment, and background stars. The reasons for the abundance variations seen for some ice features are further explored through analysis of spatial differences and correlation studies. Further, new data are used to test previous hypotheses about the carrier(s) of the XCN band. The results are discussed in \S \ref{sec:disc} with respect to different ice formation scenarios and ice chemistry in low-mass versus high-mass star-forming regions. 

\section{Sample and ice features\label{sec:obs}}

\subsection{Sample}

\noindent \textit{Spitzer}-IRS spectra were obtained for $\sim$50
low-mass protostars with ice features as part of the $c2d$ Legacy
program (PIDs 172 and 179), a dedicated open time program (PID 20604)
and a Guaranteed Time Observations (GTO) program (PI Houck). All
sources were included in Papers II-III on CO$_2$
and CH$_4$, while the $c2d$ sources alone were
investigated in Paper I and IV. 

The {\it Spitzer} spectra were complemented by ground-based Keck/NIRSPEC \citep{McLean98} and VLT/ISAAC \citep{Moorwood97} L-band spectra of the H$_2$O ice 3~$\mu$m feature. In addition, the combined sample partly overlaps with a 3--5~$\mu$m VLT survey of
CO and XCN ice toward low-mass protostars
\citep{Pontoppidan03,vanBroekhuizen05} and additional CO observations were obtained with Keck/NIRSPEC (Paper II). The VLT survey contains many
sources in  Ophiuchus, for which the \textit{Spitzer} data are taken from
GTO programs.  These sources are key for a comprehensive
investigation of how the XCN complex, composed of OCN$^-$ and
potentially a second carrier, relates to other ices. The reduction of
the spectra to obtain abundances are described in the individual {\it
 c2d} papers and the additional GTO sources have been reduced using
an identical procedure to those presented in Papers I and IV.

The low-mass sample is complemented with 9 ISO sources, representative
of high-mass protostars, from \citet{Gibb04} which were re-analyzed in
Papers I-III to ensure that the low- and high-mass ice abundances are
consistently derived using the same component analysis and band
strengths. The sources were not included in Paper IV, and NH$_3$ abundances are therefore taken from \citet{Gibb04}. In summary, our sample consists of 63 YSOs, that have been analyzed in a homogenous way. When compared with ice abundances toward background stars from \citet{Knez05} and \citet{Boogert11}, this sample spans all evolutionary stages from molecular clouds to disks and six orders of magnitude in stellar luminosities and a range of star-forming environments.

\subsection{Ice features}

\begin{figure}
\centering
\epsscale{1.0}
\plotone{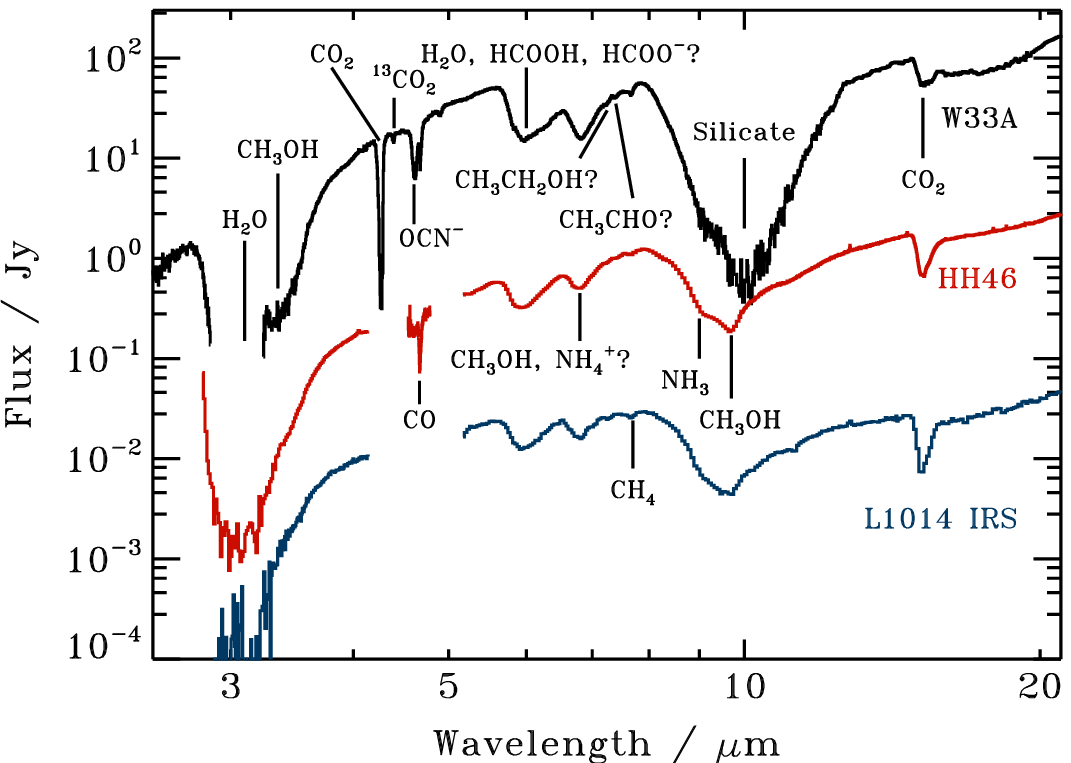}
\caption{Ice spectra toward the protostars W33A (10$^5$ L$_{\odot}$), HH46 (12 L$_{\odot}$) and L1014 IRS (0.09 L$_{\odot}$) \citep[][Paper I]{Gibb00,Boogert04b}. The 3~$\mu$m portions of the spectra have been binned to increase the S/N. \label{fig0}}
\end{figure}

Figure \ref{fig0} shows the main ice features seen toward background
stars and protostars with the identifications of different bands to
H$_2$O, CO, CO$_2$, CH$_3$OH, NH$_3$, CH$_4$ and OCN$^-$ ice
marked. H$_2$O column densities are derived from the 3 $\mu$m band
whenever such data are available; otherwise the librational band (10--30~$\mu$m)
is used. CH$_3$OH and NH$_3$ abundances are generally derived from the
9--10~$\mu$m bands. OCN$^-$ abundances are derived from one of the two
components (centered at 2165 cm$^{-1}$) that together make up the XCN
band (the total band centered at 4.62 $\mu$m next to the CO
feature). The OCN$^-$ identification is based on laboratory
spectra. The abundance of the second XCN carrier (associated with the
`2175 cm$^{-1}$' band) is derived assuming the OCN$^-$ band
strength. 

The CO and CO$_2$ spectral features can be decomposed into a number of
components corresponding to CO and CO$_2$ in different ice mixtures
\citep[][Paper II]{Pontoppidan03}.  CO is decomposed into three
components corresponding to pure CO ice, CO mixed with CO$_2$
(CO:CO$_2$) and CO mixed with H$_2$O ice (CO:H$_2$O). CO$_2$ is
similarly decomposed into four components corresponding to pure CO$_2$
ice, CO$_2$ mixed with CO (CO$_2$:CO), CO$_2$ mixed with H$_2$O ice
(CO$_2$:H$_2$O) and a shoulder which has been associated with CO$_2$
mixed with CH$_3$OH \citep{Dartois99}.

More tentative identifications of spectral bands to HCOOH, NH$_4^+$,
CH$_3$CH$_2$OH and CH$_3$CHO are also marked in Fig. \ref{fig0}. Most
of these tentative identifications are found in a complex band between
5 and 7 $\mu$m. To constrain the carriers of this band it was
decomposed into five different components (C1--C5) after subtraction
of the contribution from H$_2$O ice in Paper I. C1 and C2 make up the
6 $\mu$m band, C3 and C4 the 6.8 $\mu$m band and C5 is a broad,
underlying feature that covers the entire 5--7 $\mu$m region. From comparison with laboratory ice spectra, C1 has been identified with HCOOH and H$_2$CO, C2 with HCOO$^-$ and NH$_3$, C3 with NH$_4^+$ and CH$_3$OH, C4 with NH$_4^+$ and C5 with warm H$_2$O and anions (Paper I).

\section{Results}\label{sec:res}

\noindent Figure \ref{fig0} shows the similar 3--20 $\mu$m spectra toward protostars spanning orders of magnitude in luminosities, suggestive of a partly shared ice formation history during all types of star formation.  The spectra are not identical, however, and median abundances of identified ices toward low- and high-mass protostars are presented in \S3.2. Ice abundance variations toward low-mass protostars are further investigated in \S3.3 to constrain which ice form together with H$_2$O and which do not. In \S3.4 the low-mass sample is contrasted with ice abundance distributions toward high-mass protostars to test the significance of any differences noted in \S3.2. \S3.5 continues with a comparison of ice distributions between clouds and protostars to constrain which ices require protostellar processing to form. \S3.6. presents an analysis of spatial variations in ice abundances, both on large-scale between clouds, and on small scales within a single core. Finally, ice correlations are used in \S3.7 to provide further constraints on XCN formation, and H$_2$O ice anti-correlations are presented to investigate which ices form in competition with H$_2$O.

%This information is used together with spatial and abundance correlations and statistical tests to address which ices depend on their local environment and which ones do not, and to further identify critical environmental parameters for the ices that vary from source-to-source. %The section begins with the derivation of new ice abundances toward ten Ophiuchus sources not included in papers I,IV.

\subsection{New ice abundances in Ophiuchus}

\noindent 
To enlarge the sample of low-mass sources with available XCN data, a set of
GTO data in Ophiuchus was analyzed in the same way as the $c2d$ sources.
The five components, C1--5, from
Paper I are fitted to the 5--7 $\mu$m complex with their peak
optical depths as free parameters.  The NH$_3$ and CH$_3$OH abundances
toward the same sources are determined from their 9.0 and 9.7 $\mu$m
features, using a 4$^{\rm{th}}$ order polynomial to remove the
underlying silicate feature (Paper IV). After continuum subtraction,
column densities are derived by fitting two Gaussians to the observed
spectra around the expected band positions and using the same band
strengths as in Paper IV (Table \ref{tab2:5-8um}). The flux and optical
depth spectra are shown in Appendix A together with fitted peak
positions and band widths. 
While NH$_4^+$ is not explicitly discussed here except in the context of
the N-budget, maximum abundances for the Ophiuchus sources can be
extracted from Table \ref{tab2:5-8um} using the same conversion factor as in
Paper I. HCOOH abundances are not reported, since the previously used
feature may have a significant contribution from other carriers
(Appendix B). The presence of a feature at 7.25~$\mu$m is in itself evidence for the formation of complex ices, even though its dominating carrier (probably HCOOH or CH$_3$CH$_2$OH) is speculative.

\subsection{Median abundances}

\begin{figure}[htp]
\centering
\plotone{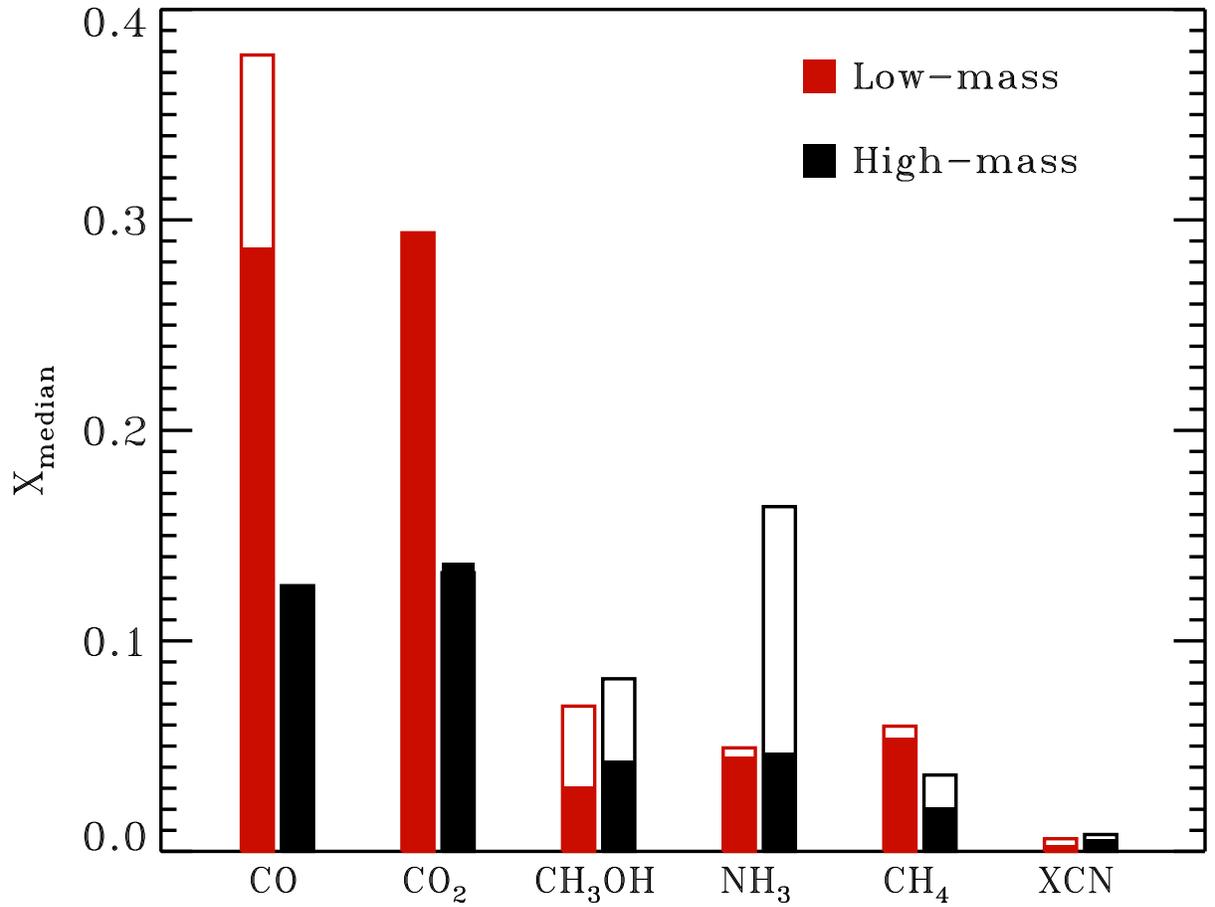}
\caption{Median abundances with respect to H$_2$O in the low- and high-mass protostellar samples (X$_{\rm median}$) derived using survival analysis that includes upper limits (filled bars) and median detected abundances (outlined bars) toward low- and high-mass protostars.\label{fig2}}
\end{figure}

\noindent Figure \ref{fig2} shows the median ice abundances with 
respect to H$_2$O ice for low- and high-mass protostars, calculated from 
Paper I-IV and from the new Ophiuchus values reported in Table \ref{tab2:5-8um}. All ice abundances are expressed with respect to H$_2$O because H$_2$O ice forms early during star formation, is the most abundant ice, and has a high sublimation point. The median values are calculated: 1) using only the detections, and 2) using the Kaplan-Meier (KM) estimate of the survival function. The latter is a non-parametric procedure that takes into account the constraints provided by upper limits. When calculating the KM estimate the detections and upper limits are ordered from low to high and the upper limits are given the values of the nearest lower detections. For example in a sample of four detections of 0.5, 1, 3 and 4 and an upper limit of $<$2, the upper limit is treated as a detection of 1. The KM estimate and how to apply statistical tests on it is reviewed by \citet{Feigelson85}. The medians calculated from the KM estimate provide more accurate descriptions of ice populations with significant upper limits as demonstrated in Fig. \ref{fig2}, where the low- and high-mass CH$_3$OH medians and the NH$_3$ and CH$_4$ median toward high-mass protostars are 40-70\% lower when taking into account the upper limits.

Table \ref{tab:median} lists the medians that best describe the data set for total ice abundances, while Table \ref{tab:median2} provides medians for all ice components together with their lower and upper quartile values. Quartiles are favored over standard deviations since most ice abundances do not follow a normal distribution (see below). For the species where the Kaplan-Meier estimate  is $>$20\% lower than the detection-based median, both are listed. CO and CO$_2$ are the most abundant ices after H$_2$O, with low-mass protostellar median abundances of 29\% with respect to H$_2$O. NH$_3$, CH$_3$OH and CH$_4$ have comparable median abundances of 3-5\%, while XCN is rare at $<$1\%. The CH$_3$OH and NH$_3$ abundance medians are similar toward low- and high-mass protostars, when the upper limits toward high-mass protostars are taken into account, while CO, CO$_2$ and CH$_4$ are more abundant and XCN is less abundant toward low-mass protostars. The low CO$_2$ abundances in this high-mass sample are consistent with recent observations of CO$_2$ ice abundances of 10--18\% toward high-mass protostars in the galactic center, but not toward high-mass YSOs in the LMC \citep{An09,Oliveira11}. Testing whether perceived differences between low- and high-mass protostars are significant requires an investigation of the ice abundance distributions within each sample of protostars, however, and this is done in \S3.4 after presenting ice abundance variations toward low-mass protostars in \S3.3.

\subsection{Ice abundance variations}

\begin{figure}[htp]
\centering
\plotone{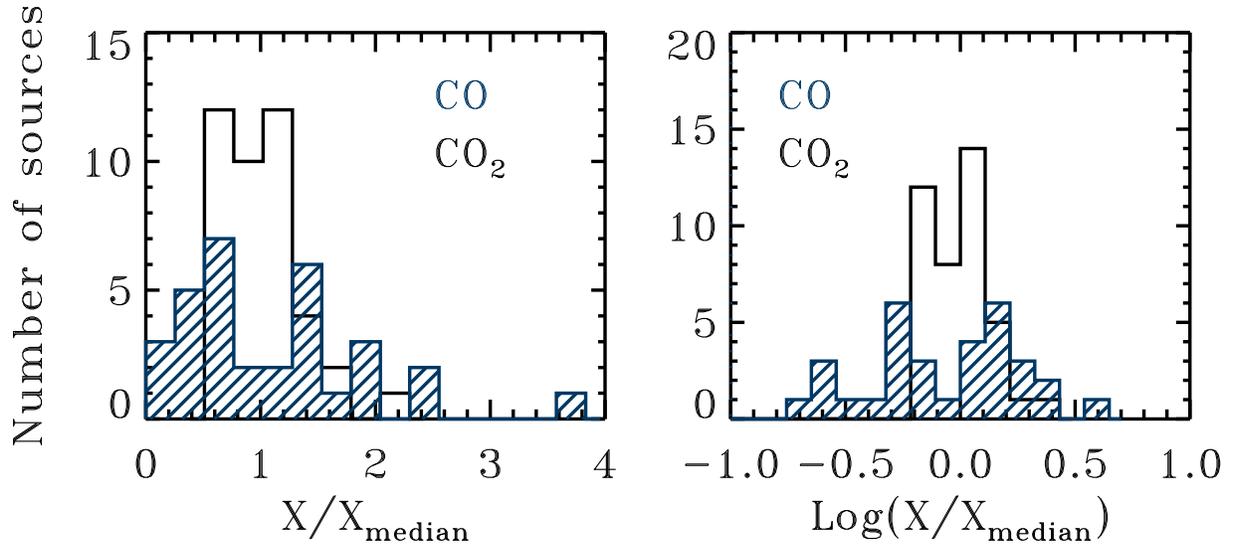}
\caption{Histograms of the CO (blue line-filled) and CO$_2$ (black contours) abundances toward low-mass protostars with respect to the median abundances. The left panel shows the skewed abundance distributions of both molecules and the right panel the more symmetric log-normalized histograms. \label{fig3}}
\end{figure}

\noindent Ice abundance variations between different sources depend on how sensitive the ice formation and destruction pathways are to the local environment.  Investigating protostellar abundance variations thus puts constraints on when and where different ices form (variations are not dominated by reported abundance uncertainties from fitting the observed spectra). 
Figure \ref{fig3} demonstrates that some ice abundance distributions are skewed with a high-abundance tail. All abundance histograms are therefore log-transformed and then centered on the median detected low-mass protostellar ice abundance, with bin sizes proportional to the low-mass abundance variances. For ices where the median is unaffected or barely affected ($<$20\%) by including upper limits, the detections alone are shown. For ice abundances that are better constrained by including upper limits (CO, CH$_3$OH, XCN, OCN$^-$, 2175 cm$^{-1}$, C2 and C5 toward low-mass protostars and CH$_3$OH, NH$_3$, CH$_4$, XCN, CO:CO$_2$ and C5 toward high-mass protostars), the distributions include 3-$\sigma$ upper limits. Appendix C contains figures that compare the distributions with and without upper limits included. Differences between low- and high-mass protostellar abundances are discussed further below, while this section focuses on the larger low-mass sample. 

Since ice abundances are with respect to H$_2$O, a small variation of a species indicates co-formation with H$_2$O, while large ice abundance variations are indicative of different formation and/or destruction dependencies than H$_2$O ice. Figure \ref{fig4} shows that the total CO$_2$, CH$_4$ and NH$_3$ abundance distributions are narrow, while CO, OCN$^-$, the 2175 cm$^{-1}$ XCN component and CH$_3$OH have broader distributions with order of magnitude abundance variations between different sources. Figure \ref{fig5} shows that the narrow CO$_2$ distribution is due to CO$_2$:H$_2$O ice, while all other CO$_2$ and CO component distributions are broad. The pure CO and CO$_2$ ice components have the broadest distributions, consistent with their expected dependence on the protostellar envelope temperature for ice evaporation and segregation. The 5--7 $\mu$m components, C1--5, span the full range of variations observed among the identified ice components, including the previously noted narrow C3 distribution (Paper I). 

Protostellar ice heating was explored in Paper I and II as explanations for observed abundance variations, using pure CO/CO:H$_2$O and pure CO$_2$/CO$_2$:H$_2$O as diagnostics of the amount of moderate ice heating, and H$_2$O/silicate as a probe of complete ice sublimation. Correlations between H$_2$O/silicate and the different C1-5 components are extensively discussed in Paper I and by \citet{Boogert11} for protostars and background stars. The main conclusion was that C3/C4 does depend on ice sublimation, supporting its identification with low- and high-temperature NH$_4^+$ salts.

Moderate ice heating is predicted to reduce the abundances of volatile ices (decreasing the pure CO abundance), segregate previous ice mixtures (resulting in pure CO$_2$), drive ice crystallization, activate acid-base chemistry and cause diffusion of ice radicals and thus the formation of more complex species. None of the ice abundances in this study are however correlated with either ice temperature tracer (not shown) except for the previously found C5/H$_2$O correlation with pure CO/CO:H$_2$O (Paper I). Moderate (transient) ice heating alone seems to play a minor role in simple ice formation. This result does not exclude that the combined differences in temperature and UV radiation around low- and high-mass protostars results in different ice chemistries. Whether this is the case is the topic of next section. 

\subsection{Low- versus high-mass protostars} 

\begin{figure}[htp]
\centering
\plotone{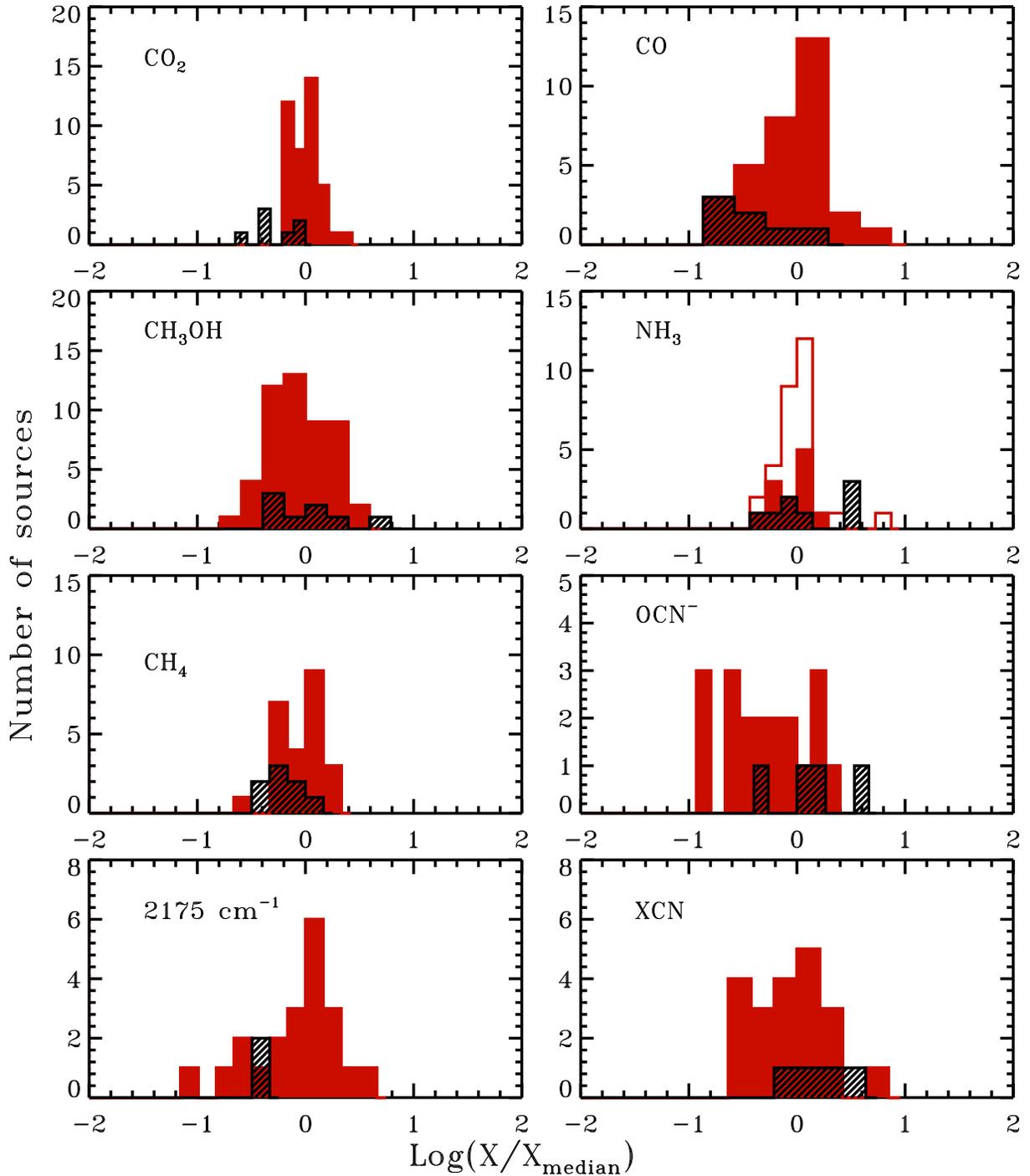}
\caption{Histograms of total ice abundances toward low-mass protostars (red) and high-mass protostars (black). For low-mass NH$_3$ abundances, the solid filled histograms are from Paper IV using a silicate template to extract NH$_3$ and the contoured histograms are from Paper IV and Table 1 when using a polynomial to fit the local continuum around the NH$_3$ feature. \label{fig4}}
\end{figure}

\begin{figure}[htp]
\centering
\plotone{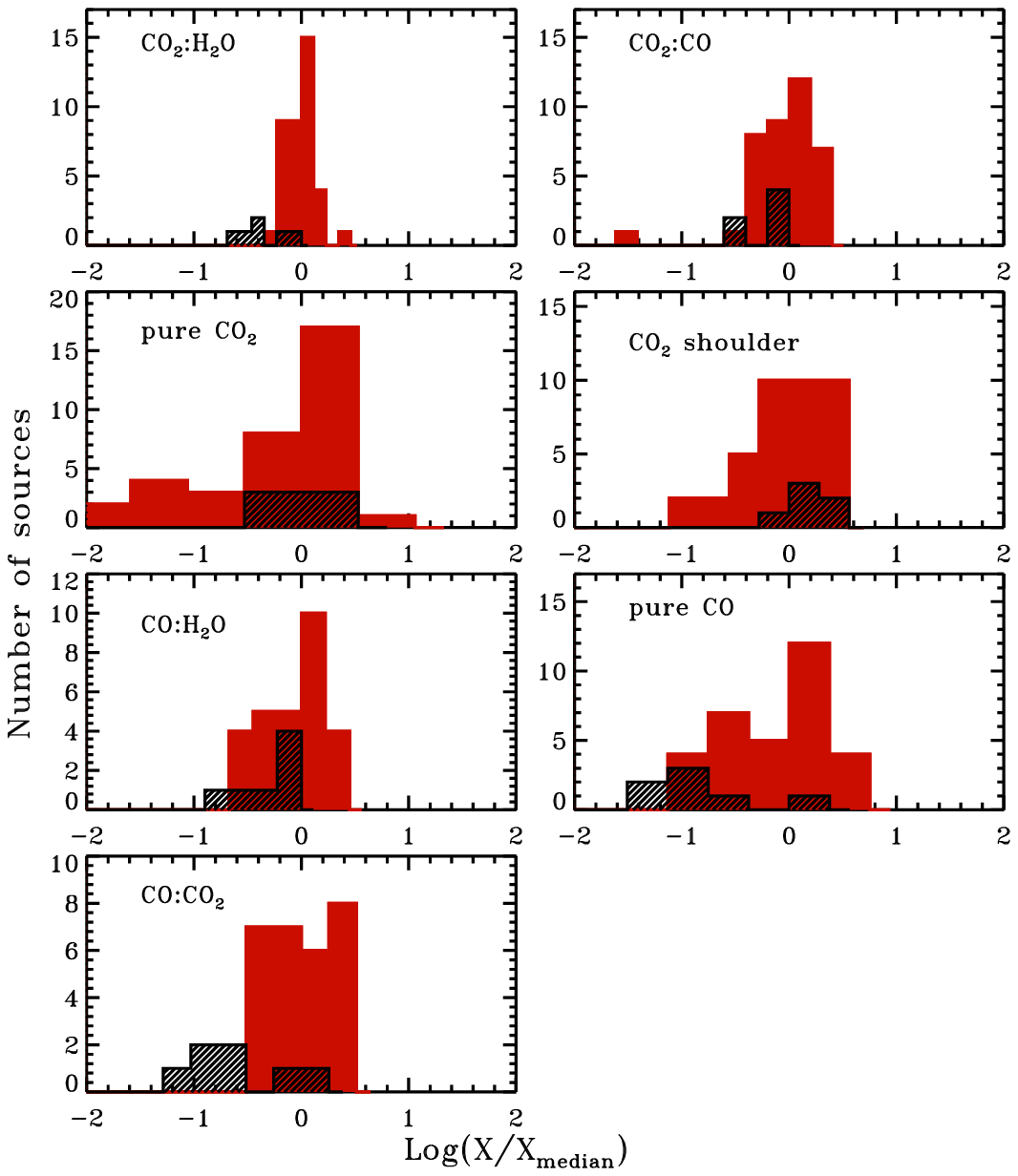}
\caption{Histograms of CO and CO$_2$ components, otherwise as in Fig. 4. \label{fig5}}
\end{figure}

\begin{figure}[htp]
\centering
\plotone{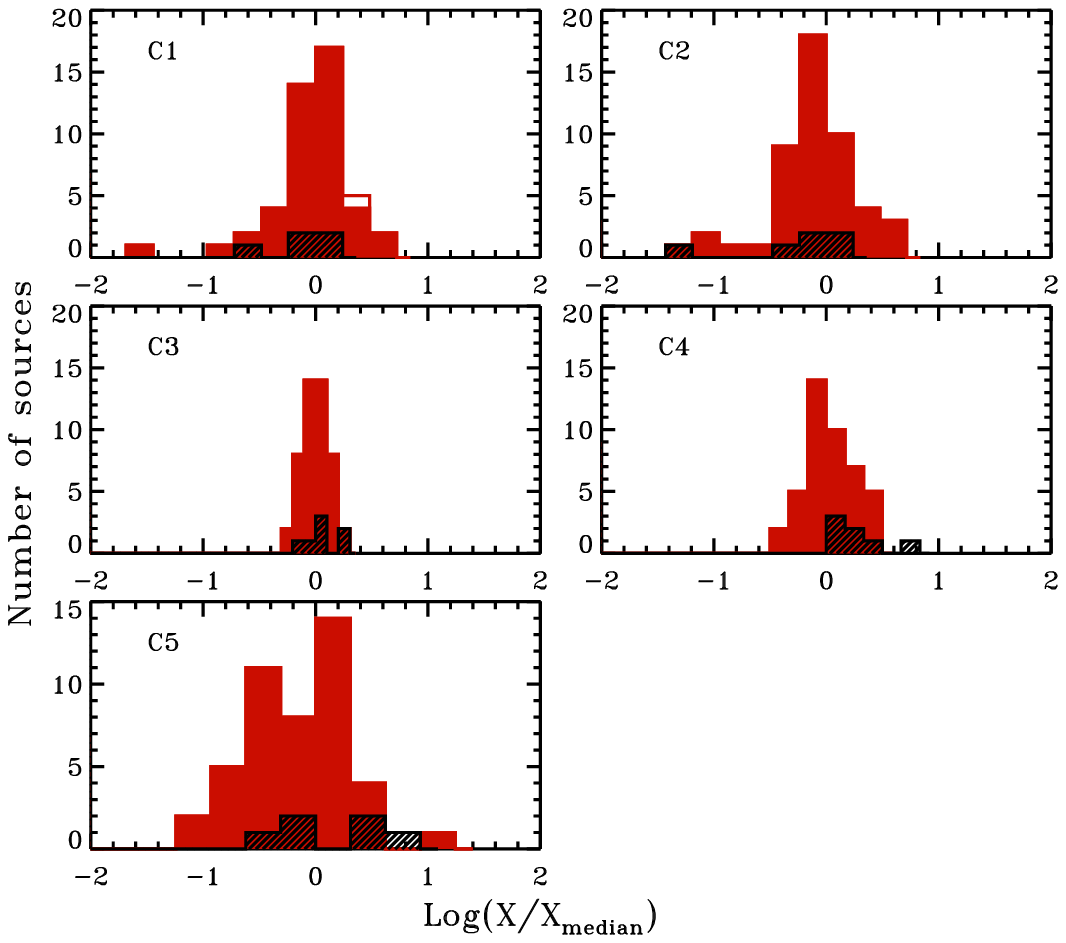}
\caption{Histograms of 5-7 $\mu$m components, otherwise as in Fig. 4. \label{fig6}}
\end{figure}

Figures \ref{fig4}--\ref{fig6} show that many ices are similarly distributed toward low- and high-mass protostars, but as noted in \S3.2, the total CO$_2$, CO and CH$_4$ abundances are lower toward the high-mass sources. 
Figure \ref{fig5} reveals that the difference in CO$_2$ abundances between low-mass and high-mass stars is due to a difference in the CO$_2$:H$_2$O ice component, which typically dominate the spectra toward low-mass sources (Paper II). All CO components are centered around a lower value toward high-mass protostars compared to low-mass protostars. Among the C1--5 components, C4 seems more abundant toward high-mass protostars. 

The significance of the visual differences in Figs.  \ref{fig2},\ref{fig4}--\ref{fig6} and other proposed differences between low-mass and high-mass protostars can be evaluated statistically. Kruskal-Wallis one-way analysis of variance by ranks is a non-parametric method to test the equality of medians in two or more groups of observations \citep{Kruskal52}. The test statistic does not assume normal distributions and is given by 

\begin{equation}
K=(N-1)\frac{\Sigma^g_{i=1}n_i(\bar{r}_i-\bar{r})^2}{\Sigma^g_{i=1}\Sigma_{j=1}^{n_i}(\bar{r}_{ij}-\bar{r})^2},
\end{equation}

\noindent where $g$ is the number of groups, $n_i$ is the number of observations (ice abundances) in group $i$ (low or high-mass protostars), $r_{ij}$ is the rank of observation $j$ in group $i$ (here the rank of a source in terms of the investigated ice abundance), $N$ is the total number of observations across all groups, $\bar{r}_i$ is the mean rank of group $i$ and $\bar{r}$ is the mean rank in the whole population. The null hypothesis of equal medians is rejected when $K\geq\chi^2_{\alpha:g-1}$ where $\alpha$ is the significance level and $g-1$ the degrees of freedom for the $\chi^2$ statistic. The $\chi^2_{\alpha:g-1}$ values can be looked up in tables and are also part of most statistic packages such as $R$.

Applying the test to the low-mass versus high-mass detected ice abundances shows that there is a significant difference (at the 95\% level) between the two samples for CO$_2$, CO, CO$_2$:H$_2$O, pure CO and C4 ice abundances. Differences in C5, CH$_3$OH or OCN$^-$, which have been suggested to be more efficiently formed toward high-mass protostars, are not statistically significant. To check whether including upper limits affect these results, a similar test (the logrank test) was run in $R$ using the previously calculated Keplar-Meier estimates and the function {\it survdiff}. The results are consistent with the Kruskal-Wallis test, except that CH$_4$ is different in low -and high-mass stars at the 98\% confidence level when including upper limits.

\begin{figure}[htp]
\centering
\plotone{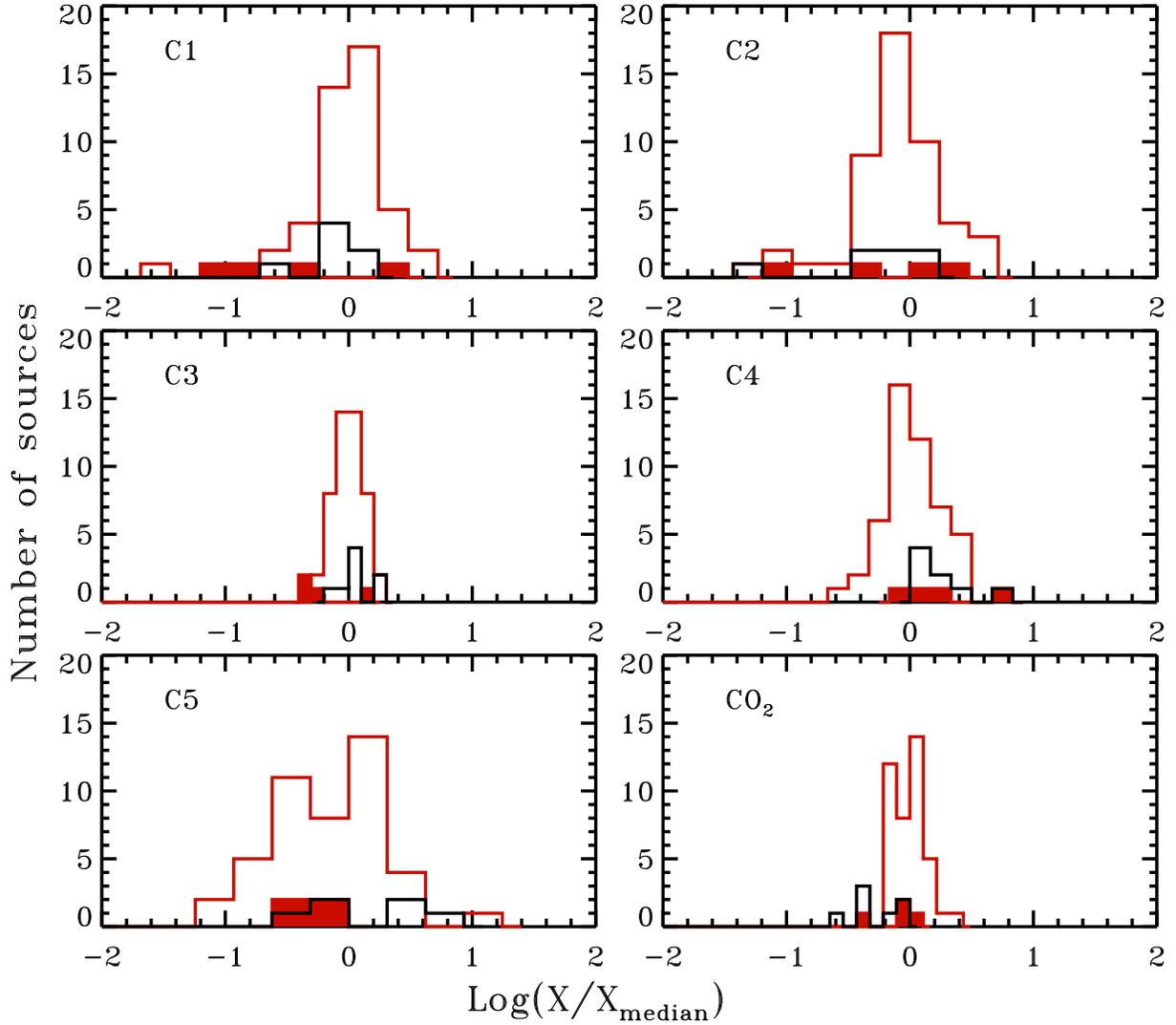}
\caption{Histograms of CO$_2$ and the C1-C5 components (including upper limits) toward low-mass protostars (red contours), high-mass protostars (black contours) and the four IC 1396A sources (red solid). \label{fig7a}}
\end{figure}

In the above analysis, all low-mass protostars are situated in star-forming regions without any nearby massive stars. It is therefore difficult to disentangle which differences are intrinsic to the low- versus high-mass star-forming processes and which differences depend on the local radiation environment.  IC 1396A, known as the Elephant Trunk Nebula, is a dense globule, excited by the 4 Myr old O6 star HD 206267. Ice observations -- H$_2$O, CO$_2$ and C1-5 -- toward four low-mass protostars are presented by \citet{Reach09}. The C3/C4 ratio toward these protostars resembled the ratio found toward high-mass protostars, suggesting that the ice evolution around low-mass protostars depends on the external radiation environment. Figure \ref{fig7a} shows a histogram comparison for the CO$_2$ and C1-5 abundances toward the low-mass, high-mass and IC 1396A samples. Visually the IC 1396A C3 and C4 abundances seem to better overlap with the high-mass sample, while the IC 1396A C5 upper limits and CO$_2$ detections are more consistent with the low-mass sample. None of these differences are, however, significant, nor is the difference in C3/C4 ratio, likely due to the small IC 1396A sample size. %This lack of a strong effect supports a scenario where most ices form quiescently in the prestellar phase.

\subsection{Protostars versus background stars}

\noindent Comparisons between protostars and background stars provide direct limits on which species form before the protostar turns on and starts to heat and irradiate its surroundings. Tables \ref{tab:median} and \ref{tab:median2} lists the medians\footnote[1]{The background star medians are based on detections alone, since applying survival analysis shows that including upper limits does not constrain the medians further. } of species detected toward background stars in Serpens and in individual cores \citep{Knez05,Boogert11}. For the isolated core sample we only use those targets with well known H$_2$O column densities from 3 $\mu$m observations. The upper limits on CH$_4$, NH$_3$ and OCN$^{-}$ are generally too high to be useful and there are only a handful of CO ice observations. C1-5, CO$_2$ and CH$_3$OH are, however, detected in a larger number of sources and can be used to compare cloud and protostellar ices.

\begin{figure}[htp]
\centering
\plotone{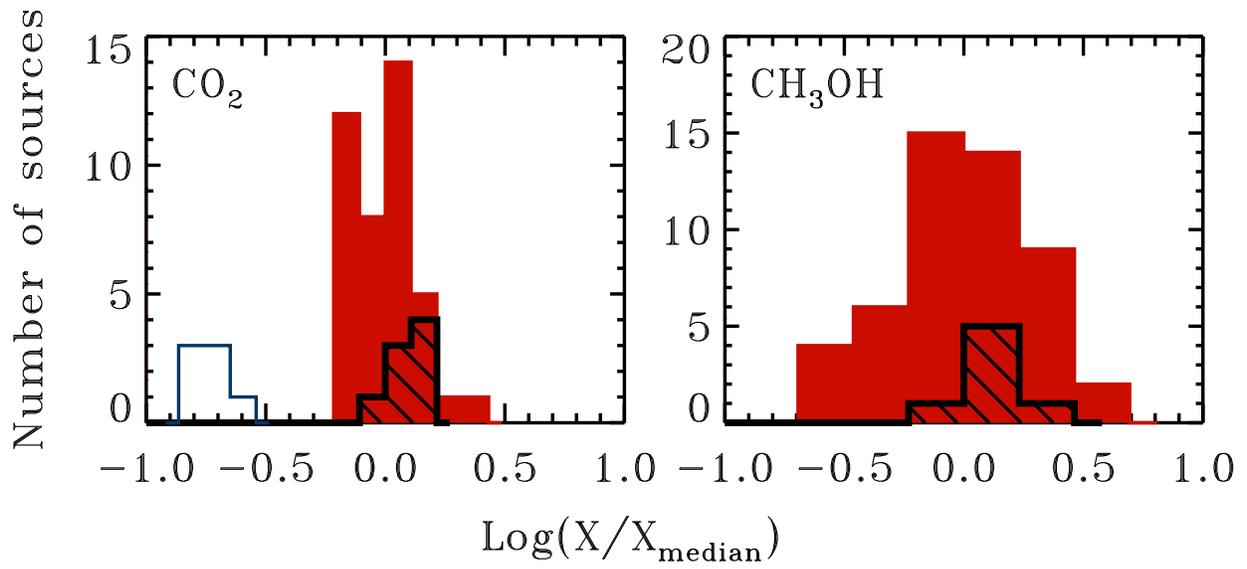}
\caption{Histograms of CO$_2$ and CH$_3$OH toward low-mass protostars (red fill), our background star sample (black line-fill) and Taurus background stars (blue contours). \label{fig7}}
\end{figure}

Previous ice studies toward Taurus revealed a lower CO$_2$ ice fraction in clouds compared to in low-mass protostellar envelopes \citep[][Paper II]{Whittet07} which together with laboratory studies have encouraged hypotheses of CO$_2$ formation from energetic processing of CO ice around protostars \citep[e.g., ][]{Ioppolo09}. Similarly, CH$_3$OH ice has been proposed to be a product of protostellar UV ice-processing \citep{Gibb04}. Figure \ref{fig7}, using data from \citet{Boogert11}, shows that there is no evidence for different CO$_2$ and CH$_3$OH ice abundances toward protostars and background stars. Applying the Kruski-Wallis and Keplar-Meier based tests confirm this lack of a difference in ice composition between our protostellar and cloud sources. Figure \ref{fig7} also shows the clear difference between Taurus CO$_2$ ice abundances from \citet{Whittet07} and our background star ice sample. From this difference, ice abundances toward Taurus background stars are not good templates for ice abundances toward average cloud cores.

In summary, observable ices, except for pure CO$_2$ ice and the C5 component, are present at the same abundances in the pre-stellar and protostellar phase. Most protostellar ice abundance variations reported in \S 3.3 must then be due to differences between clouds and to different pre-stellar ice formation processes, rather than thermal and UV processing by the protostar.

\subsection{Spatial differences within the low-mass protostellar sample}

\subsubsection{Cloud-to-cloud variations}

Variations in ice formation and destruction efficiencies in the pre-stellar stage may be associated with different cloud structures and initial chemical conditions in different star-forming regions. This is investigated by dividing the low-mass protostellar sample into six groups with sources belonging to the star forming associations in Ophiuchus, Serpens, Corona Australis, Perseus, Taurus and a collection of smaller associations. Kruskal-Wallis tests show that when considered as a whole, the ice abundance medians are statistically indistinguishable between these groups. This is consistent with a Principal Component Analysis in \citet{Oberg09f}, which revealed no differences between the different clouds when simultaneously considering all ice abundances.

Cloud-to-cloud variations can also be tested for specific ice components. CO$_2$ ice abundances in clouds were shown above to be unusually low toward Taurus, while too few CO measurements toward background sources outside of Taurus exist to determine whether CO is lower than expected as well.  Testing the CO and CO$_2$ protostellar differences between the clouds may reveal whether prestellar differences are carried over into the protostellar phase. CH$_3$OH is another ice that has been observed to be spatially variable \citep{Pontoppidan04,Boogert11}. Differences in the pure CO ice component and the CO$_2$ 
shoulder (often assigned to CO$_2$:CH$_3$OH interactions) between the clouds are indeed statistically significant at a $>$98\% confidence level, while all other CO and CO$_2$ components and CH$_3$OH are not. The significant differences exist between the low abundances toward Taurus and the smaller associations on one hand, and the more 'normal' abundances toward the other clouds (Table \ref{tbl:anova}). 

The difference between Taurus and the other clouds is important since much of the current ice paradigm is based on analysis of Taurus sources \citep[e.g.][]{Whittet07}.
The low pure CO ice abundances may also explain the low CH$_3$OH ice upper limits toward Taurus, since CO freeze-out is a pre-requisite for CH$_3$OH ice formation without energetic radiation \citep{Cuppen09} and a low pure CO ice abundance may indicate that catastrophic CO freeze-out has not taken place. In summary, some CO ice abundance differences seem to be explained by large-scale differences in cloud environments, but overall most cloud and pre-stellar ice formation variability must be due to more local effects than cloud-to-cloud variation. Both from cloud and protostellar ice abundances, Taurus stands out and care should hence be taken in using Taurus trends as a basis for general conclusions on ice evolution during low-mass star formation.

\subsubsection{An ice map toward the Oph-F core \label{sec2:res_maps}}

\begin{figure}[htp]
\centering
\plotone{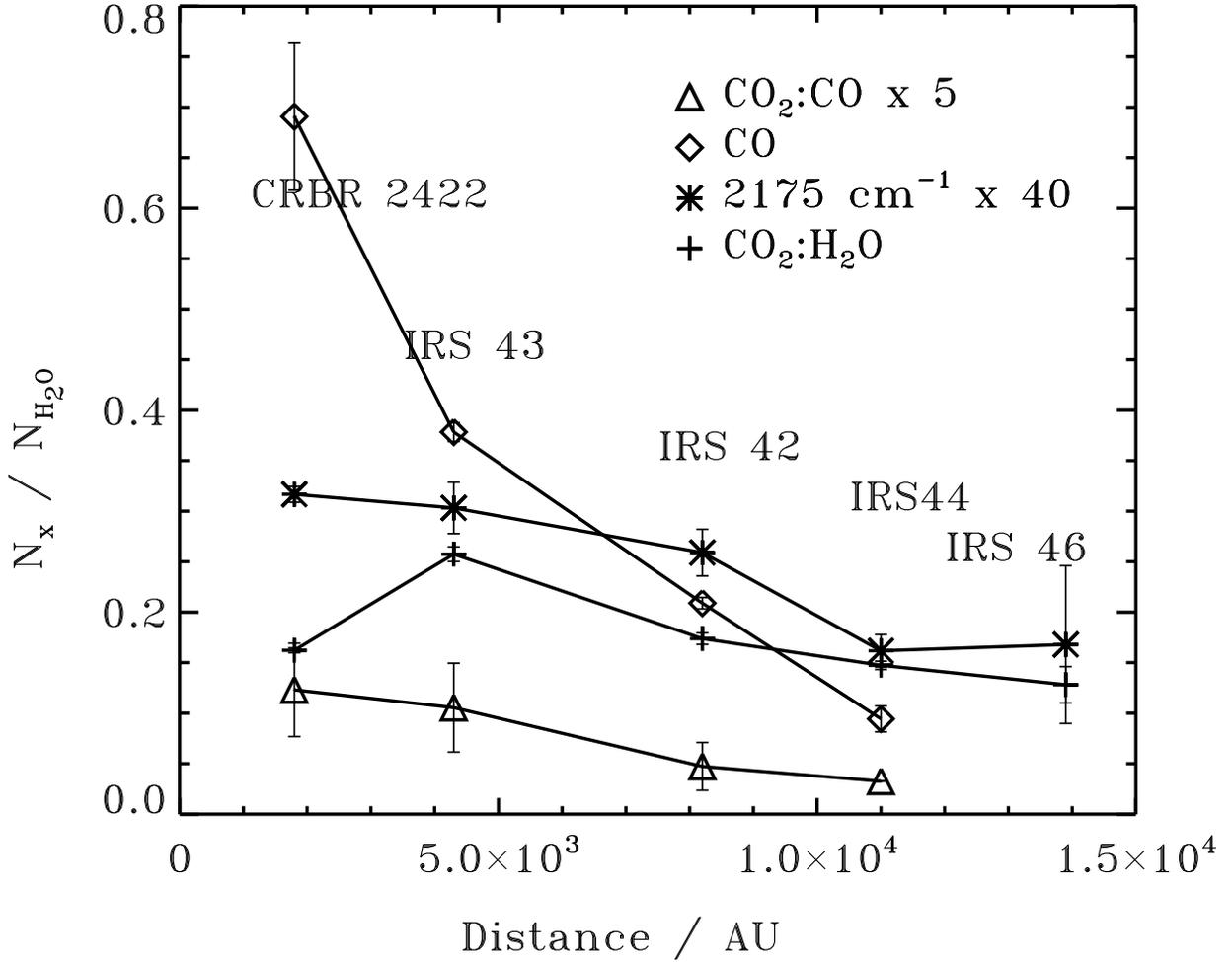}
\caption{Ice abundances at different distances from the Oph-F core with respect to H$_2$O ice. The 2175 cm$^{-1}$ abundances are scaled by 40 and the CO$_2$:CO abundances by 5 for clarity. The CO$_2$ values are taken from Paper II, rather than \citet{Pontoppidan06}, which affects especially the CO$_2$:H$_2$O curve.  \label{fig8}}
\end{figure}

\noindent Within single cloud cores, ice maps may reveal production and destruction pathways that depend on the local environment. An ice map of protostars in the Oph-F core has been used previously to show that the total CO and the CO:H$_2$O abundances decrease monotonically away from the central core \citep{Pontoppidan06}. The lines of sight probe primarily the dense quiescent core, rather than ice in the protostellar envelopes, and the spatial trends were interpreted as catastrophic freeze-out of CO in the pre-stellar stage once a certain density and temperature is reached. Figure \ref{fig8} shows that the order of magnitude increase in CO ice with respect to H$_2$O toward the core is accompanied by an increase in CO$_2$:CO. In contrast, the CO$_2$:H$_2$O ice component is almost constant across the core. The XCN band (the 2175 cm$^{-1}$ feature -- OCN$^-$ is not detected) is the only other species that increases monotonically toward the densest part of 
Oph-F. CO$_2$:CO and the 2175 cm$^{-1}$ band thus appear directly related to CO freeze-out. No trend with distance to the core center is seen for any other ice component and is also not expected  for ices forming early during cloud formation (e.g. CH$_4$ and NH$_3$), which are independent of cloud core time scales and CO freeze-out, or of species dependent on protostellar heating (e.g. pure CO$_2$ ice) or of components with potentially multiple carriers, such as the C1-5 bands. CH$_3$OH is only detected toward one of the sources and no trend can thus be extracted. Still the Oph-F map suggests that CO freeze-out followed by a CO driven ice chemistry may account for large pre-stellar ice variations. 

\subsection{Protostellar correlation studies}

The trends found for the Oph-F core are here explored for the entire protostellar sample through correlation studies between especially CO-chemistry products and H$_2$O ice column densities,  and between XCN, and CO and CO$_2$ ice components.

\subsubsection{Dependencies on H$_2$O ice column \label{sec2:res_h2o}}

\noindent In individual cores, the H$_2$O column density has been shown to correlate well with dust extinction, assumed to trace the total dust mass and H$_2$ column \citep{Whittet88,Pontoppidan04}. This implies that the average abundance of H$_2$O ice per dust grain is constant across the core. In an arbitrary line of sight, the total H$_2$O ice column density depends on both the column density of dust grains and on the average number of monolayers of H$_2$O ice on each dust grain. If variations in the H$_2$O column density across the sample are dominated by the line of sight fraction of high-density material, ices dependent on CO freeze-out are expected to correlate with H$_2$O column density, since CO freezes out catastrophically at high densities (\S3.6.2). If instead variations in the H$_2$O column density across the sample is dominated by the H$_2$O abundance on each grain, the main observable trend should be that ices that form in competition with H$_2$O anti-correlate with the H$_2$O column density.

In theory, the dust column density can be estimated from the optical
depth of the 9.7~$\mu$m silicate feature or from the continuum extinction.
Because of a number of practical complications we do not attempt to
normalize the H$_2$O column density to the dust column, however.  First,
toward protostars, the silicate absorption can be filled in by an
unknown amount of silicate emission from the warmer parts of the
envelope.  Second, continuum extinction estimates are affected by
uncertainties in the intrinsic protostellar SEDs. Third, while the 9.7
$\mu$m band correlates well with the near-infrared color excess, the
relations vary in different environments by factors of 2 \citep{Chiar07, Boogert11}, possibly because of variations in the
grain size distribution or even variations in the grain composition \citep{vanBreemen11}. And fifth, the contribution of unrelated,
ice-less foreground dust is often unknown.

\begin{figure}[htp]
\centering
\epsscale{0.95}
\plotone{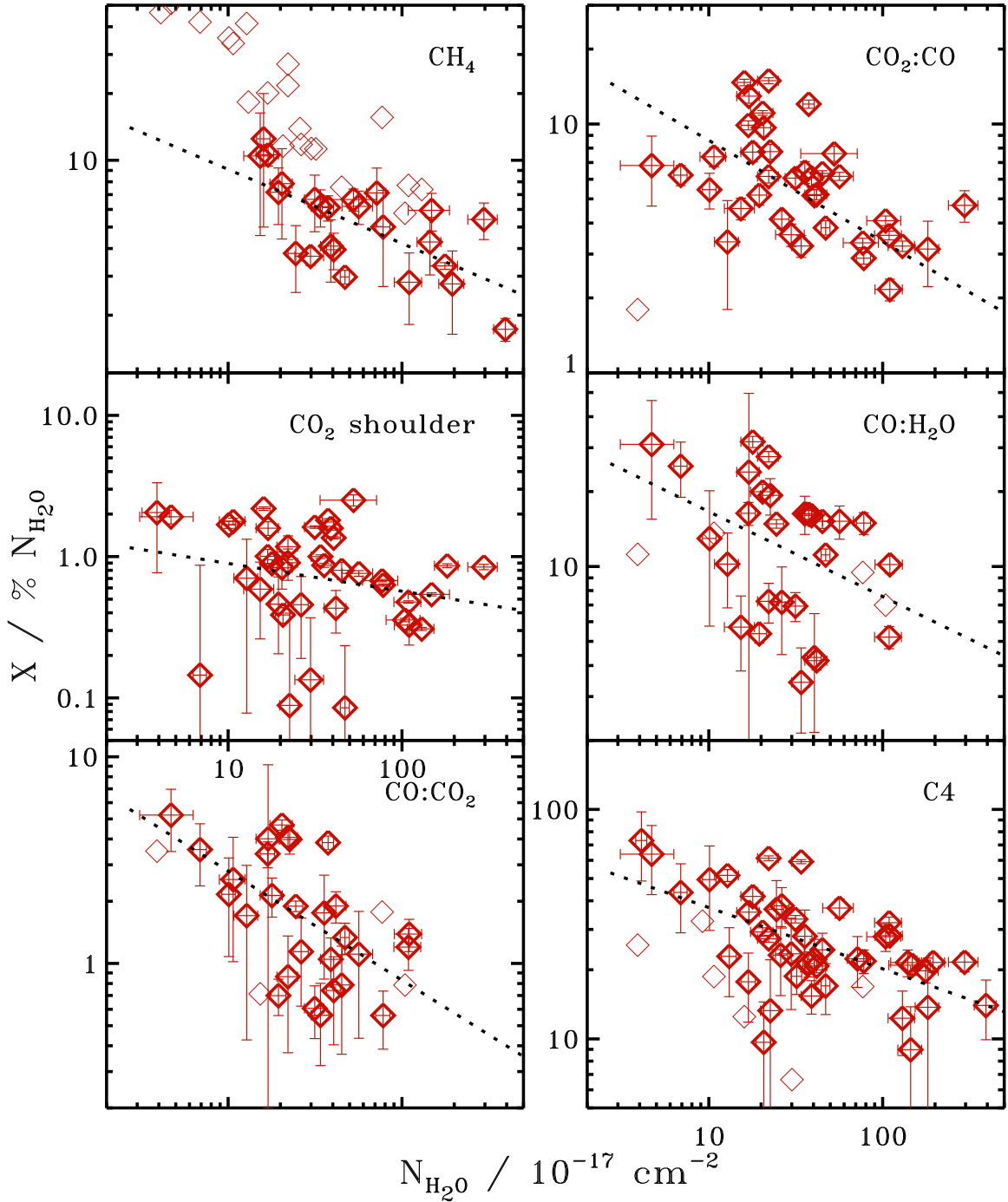}
\caption{Statistically significant correlations between ice abundances w. r. t. H$_2$O and the H$_2$O ice column density for low-mass protostars, with  upper limits plotted with thin symbols. The significance of the correlations was measured using Spearman's rank correlation test, which makes no assumptions about the type of correlation, while the dotted line shows the best log-log fit to the data to guide the eye. \label{fig8b}}
\end{figure}

There are no significant positive correlations (at the 95\% confidence level)  between H$_2$O and any ice abundance, X/H$_2$O, when applying Spearman's rank correlation test \citep{Spearman04} to the low-mass protostellar sample; rank correlation tests are more robust to outliers compared to most parametric correlation tests and make no assumptions about whether the correlation is linear or non-linear. Figure \ref{fig8b} shows that instead, six ice components -- CH$_4$, CO$_2$:CO, the CO$_2$ shoulder, CO:H$_2$O, CO:CO$_2$ and C4 -- are inversely correlated with the H$_2$O column density at the 97--99\% confidence level.  C4 has been associated with NH$_4^+$ salts and its anti-correlation with H$_2$O may be explained by an increasing salt fraction with ice sublimation because salts have higher sublimation temperatures than H$_2$O ice (Paper I). This scenario cannot, however, explain the other anti-correlations, since CH$_4$, CO and CO$_2$ are all more volatile than H$_2$O. 

The CO-related anti-correlations may be a result of competitive formation of H$_2$O and CO; the more oxygen that is bound up in H$_2$O ice, the less may be available to form gaseous CO and thus ices that depend on CO freeze-out. Still in each individual core CO ice abundances and H$_2$O ice column densities are expected to correlate because of increasing CO freeze-out  toward the densest part of the cloud core. The latter may explain some of the scatter in the plots. Finally, the CH$_4$--H$_2$O trend is mainly visible for very low H$_2$O column densities (many in Ophiuchus) and this relation may be due to more efficient CH$_4$ formation during only the earliest H$_2$O formation stage when a large fraction of carbon is still in atomic form.

\subsubsection{XCN correlations}

\begin{figure}[htp]
\centering
\epsscale{1.0}
\plotone{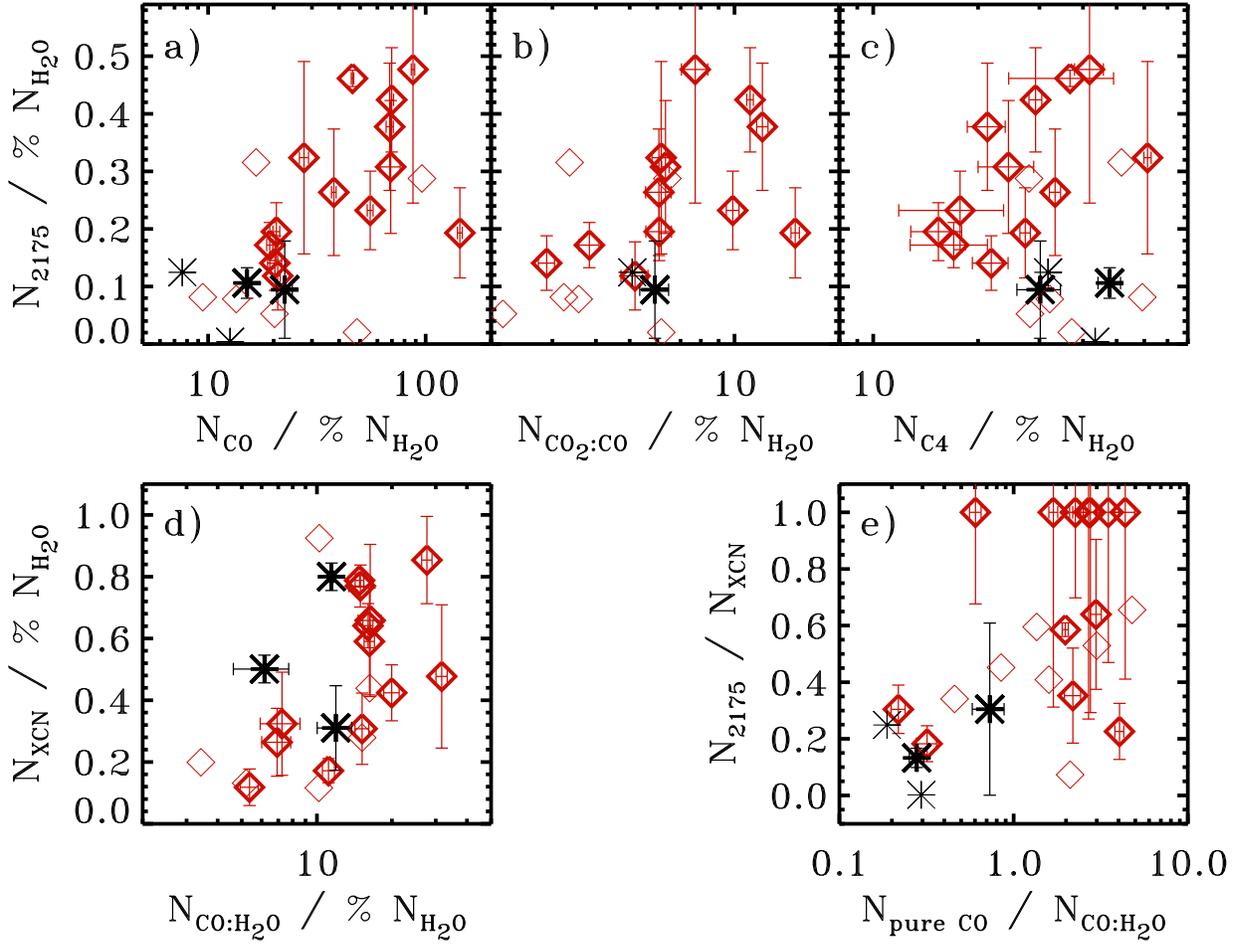}
\caption{Panels a--d show correlations between XCN and CO ice components for low-mass protostars (red diamonds) with high-mass protstellar abundances shown for comparison (black stars). Panel e shows the relationship between and ice heating tracer and the two XCN components. Thin symbols denote upper limits. \label{fig9}}
\end{figure}

\noindent As stated in \S1, the XCN band is composed of two components, peaking at 2165 cm$^{-1}$ (OCN$^-$) and at 2175 cm$^{-1}$. Toward the Oph-F core, the 2175 cm$^{-1}$ component is spatially correlated with CO, CO$_2$:CO and CO:H$_2$O. These correlations were explored in the whole low-mass sample and Fig. \ref{fig9} shows that the 2175 cm$^{-1}$ component is significantly correlated with CO and CO$_2$:CO (95\% confidence with  Spearman's rank correlation test)\footnote[2]{The Spearman's rank correlation is not yet implemented for censored data in $R$ and the test was therefore not repeated including the constraints of the upper limit. Figure~\ref{fig9} shows such an inclusion would only strongly affect the C4 correlation.} and the correlation is even stronger ($>$99\%) when the high-mass sources are added. The complete XCN band is not correlated with either of these components, but it is significantly correlated with the less volatile CO:H$_2$O component (95\% level). Panel e, finally confirms previous claims in \citet{vanBroekhuizen04} that the relative importance of the OCN$^-$ feature and the 2175 cm$^{-1}$ feature depend on ice processing, measured here by the pure CO / CO:H$_2$O ice ratio.

These correlations are indicative of a CO-related, single carrier of the entire XCN band with a spectral profile that depends on the environment. The simplest explanation is that OCN$^-$ is responsible for the entire XCN band (i.e. both the 'OCN$^-$ component' and the 2175 cm$^{-1}$ component) and that the two components are due to OCN$^-$ in a volatile (CO-rich) ice toward some sources, and in a different ice mixture in warm environments. OCN$^-$ can form at 15~K from HNCO in the presence of strong bases \citep{Raunier03,vanBroekhuizen04}. Consistent with the 2175 cm$^{-1}$ identification with OCN$^-$,  this component seems to be correlated (95\% level) with C4 (one of the two bands ascribed to the base NH$_4^+$ in Paper I) when only considering low-mass protostellar detections.  The correlation is not significant,  however, when including the high-mass sample or the significant upper limits and must therefore be considered as tentative.

\section{Discussion} \label{sec:disc}

\subsection{C-, O-, and N-budget}

The amount of C, O and N that are typically bound up in ice mantles is important for the  life cycle of the elements in the interstellar medium. The total (refractory + volatile) C, O and N abundances in the solar neighborhood are 2.1, 5.8 and 0.58$\times10^{-4}$ per hydrogen nucleus, respectively \citep{Przybilla08}. The fractional C, O and N abundance in ices with respect to these total C, N and O abundances are calculated from the presented ice abundances with respect to H$_2$O ice together with literature values of the H$_2$O abundance with respect to hydrogen. \citet{Pontoppidan04} and \citet{Boogert04} modeled the H$_2$O abundance toward four of the low-mass protostars in the sample and found $n_{\rm H_2O}=4.9\pm0.7\times10^{-5}n_{\rm H}$. Toward high-mass protostars the H$_2$O abundance can be estimated using the relations A$_{\rm V}/\tau_{\rm 9.7}=9$ and N$_{\rm H}/A_{\rm V}\sim1.9\times10^{21}$  cm$^{-2}$ mag$^{-1}$ \citep{Roche85,Bohlin78}.  It is important to note that this method only provides a crude estimate since the silicate feature is affected both by emission and grain growth; a recent grain model result in a conversion factor that is 30\% lower for dense clouds \citep{Evans09}. Using the values in \citet{Gibb04} for W33A, NGC7538~IRS~9, Mon~R2~IRS~3 and S140~IRS~1 (covering the range of observed ice processing) results in a H$_2$O abundance of $(5.0\pm1.9)\times10^{-5}n_{\rm H}$. A mean H$_2$O abundance of $5\times10^{-5}n_{\rm H}$ (i.e. 10$^{-4}$ n$_{\rm H_2}$) is therefore assumed for both low and high-mass protostars.

The O abundance in the ice is calculated from $(x_{\rm H_2O}+x_{\rm CO}+2\times x_{\rm CO_2}+x_{\rm CH_3OH} + x_{\rm XCN})\times n_{\rm H_2O}$,  the C abundance from $(x_{\rm CO}+x_{\rm CO_2}+x_{\rm CH_3OH}+x_{\rm CH_4}+ x_{\rm XCN})\times n_{\rm H_2O}$ and the N abundance from $(x_{\rm NH_3}+x_{\rm XCN}+x_{\rm NH_4^+})\times n_{\rm H_2O}$, where $x_y$ is the abundance of $y$ with respect to H$_2$O. The median upper limit on  NH$_4^+$ is calculated from the total optical depth of the C3 and C4 components, using the FWHM of 0.195 and 0.292 $\mu$m, respectively from Paper I and a band strength of $4.4\times10^{-17}$ cm from \citet{Schutte03}. 

The median percentages of C, O and N bound up in ices, calculated from the ice abundances in Table 2, are listed in Table \ref{tab:con}. On average 15\% of the total (refractory + volatile) C, 16\% of the O and 10\% of the N reside in ices toward low-mass protostars. Toward high-mass protostars the same calculation yields 12\% of the O, 8\% of the C and up to 12\% of the N atoms.  The differences in the amount of C and O  are due to the low CO  and CO$_2$ abundances toward high-mass protostars.

Table \ref{tab:con} also lists the amount of volatile or non-refractory amounts of C and O that are in ices. In the solar neighborhood the non-refractory
O abundance is $3.2\times10^{-4}$ per N$_{\rm H}$ \citep{Meyer98} because a large fraction of O is bound up in silicates and is therefore not available for ice formation
\citep[e.g][]{Whittet10}. Out
of this available O, 34\% is bound up in ices toward low-mass
protostars, and 25\% toward high-mass stars. The fraction of C in refractory material is less well constrained, but assuming a non-refractory C abundance of $\sim1\times10^{-4}$ \citep{Weingartner01} results in 27\% of the available C found in ices toward low-mass protostars and 14\% toward high-mass protostars.

The above numbers all relate to the median ice abundances in the sample. The extreme ice sources may, however, provide more information on whether ice formation is limited by the elemental abundances. IRS 51 and NGC 7538 IRS~9 are calculated to be the most ice-rich sources in the sample, when still assuming a H$_2$O ice abundance of $5\times10^{-5}$ with respect to $n_{\rm H}$. Especially in the case of IRS 51, the amount of O and C found in the ices approaches the abundances of non-refractory O and C (Table \ref{tab:con}). 

It is important to note that additional O and N may
be hidden in the ice in the form of O$_2$ and N$_2$. Direct constraints on O$_2$ ice abundances are high, 50--100\% with respect to CO ice, and constraints from O$_2$ interactions with CO ice are of a similar order \citep{Ehrenfreund92, Vandenbussche99, Pontoppidan03}. O$_2$ ice is, however, unlikely to be abundant from the low upper limits on O$_2$ in the gas-phase; it is orders of magnitude less abundant than CO and since the two molecules have similar freeze-out and desorption properties it is unlikely that O$_2$ is differently partitioned between ice and gas compared to CO \citep{Goldsmith00,Pagani03,Fuchs06}. 

\subsection{Ice evolution: early and late pre-stellar ice formation}

\begin{figure}[htp]
\centering
%\epsscale{1.5}
\plotone{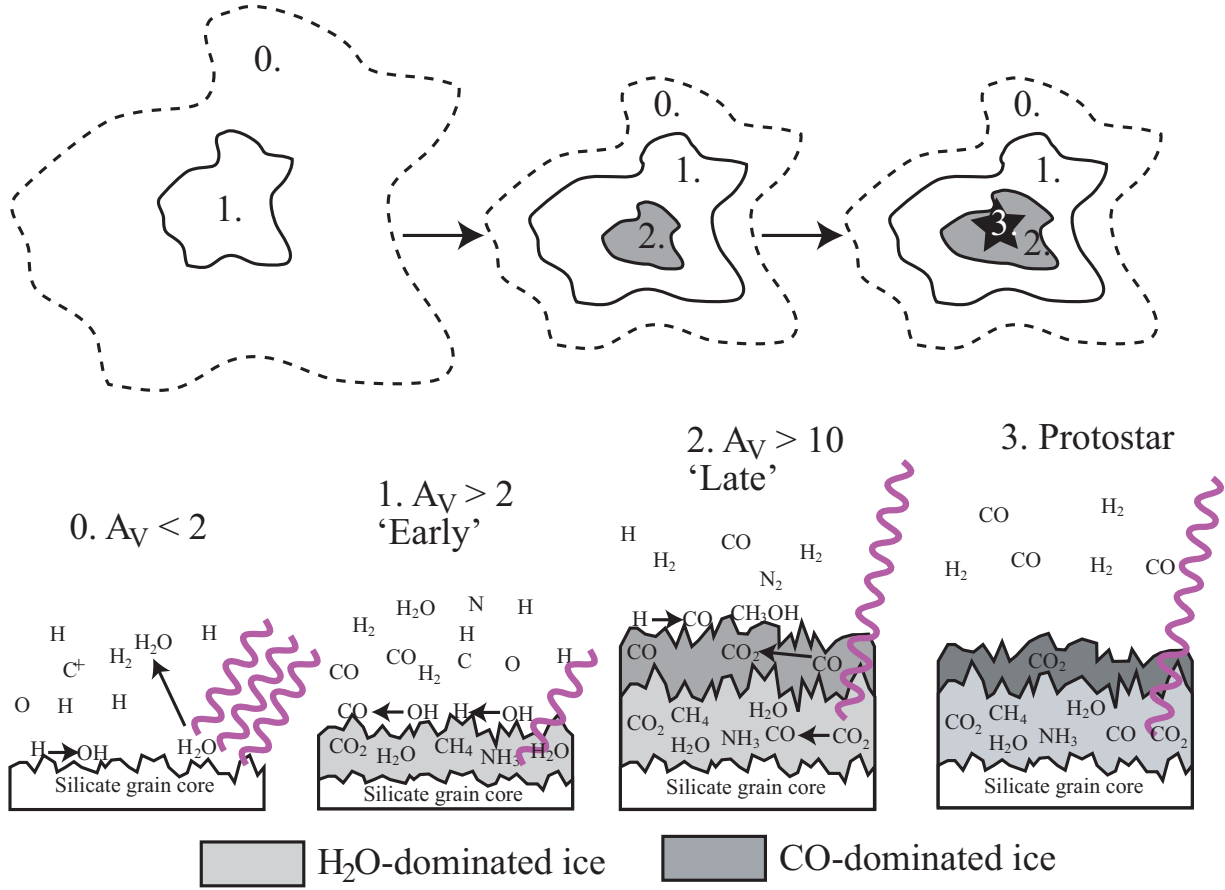}
\caption{The key ice reactions during the different stages of cloud and star formation, where 0. corresponds to cloud edges where UV photons can penetrate, 1. the early stages of dense cloud formation, 2. the later formation of cloud cores and 3. the protostellar envelope. Ice formation begins in 1. with hydrogenation of atoms resulting in a H$_2$O dominated ice with a high CO$_2$ concentration (CO$_2$:H$_2$O). At later times CO freezes out catastrophically, resulting in a second layer where CO$_2$ formation continues (CO$_2$:CO and CO:CO$_2$) and CH$_3$OH formation begins. During all cold stages small amounts of ice chemistry products are maintained in gas-phase due to non-thermal desorption. In the protostellar stage desorption of CO starts at 20~K and H$_2$O:CO$_2$ segregation at 30~K.  \label{fig11}}
\end{figure}

Toward low- and high-mass protostars alike, the ices are dominated by H$_2$O, followed by CO$_2$ and CO (\S3.2). C and O are expected to react similarly with hydrogen. The low abundance of CH$_4$ therefore implies that most C is not in atomic form when the bulk of the H$_2$O ice is forming from O  early in cloud formation or at low extinctions (Region 1. in Fig. \ref{fig11}). A fast conversion of C into CO is also consistent with the CH$_4$/H$_2$O versus H$_2$O anti-correlation toward lines of sight with low H$_2$O column densities (\S3.7); it suggests that CH$_4$ forms more efficiently in the very beginning of H$_2$O formation compared to when the bulk of H$_2$O ice forms at slightly higher extinctions. The bulk of H$_2$O formation is instead associated with continuous formation of CO$_2$, probably from CO+OH \citep{Oba10,Ioppolo11}; the CO$_2$:H$_2$O abundance barely varies from source to source in the low-mass protostellar sample (\S3.3).  H$_2$O and CO$_2$ ice mapping confirms this scenario \citep{Bergin05}, as do the similar CO$_2$ abundances toward cloud cores and protostars (\S3.5).  Most CO that is frozen out in this phase must be converted into CO$_2$. 

From its small abundance variations, NH$_3$ is also inferred to form early with H$_2$O (\S3.3). The C3 component may be associated with this stage as well and its identification with NH$_4^+$ suggests that there exists a large amount of strong acids that can protonate NH$_3$ in cold ices \citep{vanBroekhuizen04}. Anions are notoriously difficult to observe in the ice, however, and thus gas phase observations of non-thermally evaporated ices at the edges of clouds probably offer the best opportunity to confirm their existence. It is promising that HCOOH has been detected toward translucent clouds \citep{Turner99} and a more comprehensive study focusing on gas-phase HCOOH and other acids toward cloud edges, where ice formation has began and photodesorption is efficient, is warranted. 

At some point the CO/O ratio becomes large enough that most frozen-out CO is no longer converted into CO$_2$ and H$_2$O is no longer the most efficiently formed ice. We mark this as the breaking point between early H$_2$O dominated and late CO-driven pre-stellar ice formation. The ices that form in this later stage (Region 2 in Fig. \ref{fig11}) are not expected to co-vary with H$_2$O, but rather depend on the CO-freeze-out efficiency. The latter is both temperature and density dependent and may vary greatly dependent on cloud structure, cloud collapse time scales and local radiation environment. This explains why many identified ice and ice components that are definitely present during the pre-stellar stage (e.g. CH$_3$OH) have abundances that vary by orders of magnitude with respect to H$_2$O (\S3.3,3.5) both locally in a single core and when comparing different cloud complexes. In addition, the anti-correlation between H$_2$O ice and the abundance of several CO-ice components suggest that the time available for this CO-based chemistry decreases when a longer time is spent in the H$_2$O-ice formation stage, where O is converted into H$_2$O and CO$_2$:H$_2$O ice (\S3.7).

CO$_2$:CO, CO:H$_2$O and the OCN$^-$ feature are all associated with this CO-ice chemistry (\S3.5). CO$_2$:CO correlates with CO freeze-out in the Oph-F ice map, demonstrating a second, later production pathway of CO$_2$. CO$_2$ may still form from OH+CO, but at later stages CO is more abundant than O, resulting in sparse H$_2$O formation compared to CO and thus in a CO-rich CO$_2$ ice \citep[see][for a similar discussion from a theoretical point of view]{Garrod11}.  From the correlation studies in \S3.7, OCN$^-$ seems to similarly require CO freeze-out and may form from CO+NH, followed by proton transfer.  

CH$_3$OH and H$_2$CO can form from hydrogenation of CO \citep{Cuppen09} and should also depend on CO freeze-out levels. There is no strong correlation 
between CH$_3$OH 
and CO, however. The CO to CH$_3$OH conversion efficiency thus vary in different lines of sight, probably dependent on the H/CO ratio, which may vary with density and collapse time \citep{Cuppen09} and cosmic ray flux \citep{Boogert11}. Ices are continuously exposed to cosmic rays and cosmic-ray induced UV photons \citep{Shen04} and other `late' ice features may also depend on cosmic-rays, directly or indirectly. In general, observed ice features can however be explained by an early ice chemistry phase characteristic of hydrogenation of atoms, followed by atom-addition reactions in a CO dominated ice phase. %So far there is little observational evidence for radiation ice processing in the prestellar phase, however. %This variable CH$_3$OH formation efficiency may help to explain the large variations in gas-phase CH$_3$OH abundances toward hot cores \citep{Dartois99,Bisschop07}. 

\subsection{Ice evolution: the protostellar stage}

\noindent Though much of the observed ice evolution can be explained by prestellar processes, some ice features are formed or destroyed by the protostar. The broadest ice distributions belong to pure CO$_2$ and pure CO ice (\S3.3), whose variations are best explained by protostellar heating evaporating CO ice and distilling or segregating CO$_2$-containing ices \citep[Paper II][]{Oberg09e}. This last and most source dependent ice chemistry phase is also predicted to result in complex ice formation that requires ice heating for diffusion of large radicals \citep{Garrod08,Oberg09d}. While the spectra observed to date do not allow for definitive assignments to any specific complex species, the presence of e.g. a feature at 7.25~$\mu$m belonging either to HCOOH or CH$_3$CH$_2$OH, demonstrates that some complex organic ices are forming efficiently in the pre- or protostellar stage.

Most of the observed differences between low and high-mass protostars can be explained by protostellar ice heating (\S3.4). CO is comparatively volatile and its low abundances toward high-mass sources is consistent with their higher envelope temperatures. The CO:H$_2$O contents toward low and high mass objects are comparable, consistent with experiments that show that 5--10\% of CO can be trapped in H$_2$O ice up to the H$_2$O evaporation temperature \citep{Sandford90,Collings04,Fayolle11}. 

The low CO$_2$:H$_2$O abundances toward high-mass protostars can also be explained by thermal ice processing. This component starts to segregate above 30~K resulting in a pure CO$_2$ ice component, which desorbs above 50~K. The luminosity of the star will determine the radii for ice segregation and desorption and the extent of the 30--50~K shell where pure CO$_2$ ice will be common. Toward low-luminosity stars the segregation and desorption radii will be closer to the star resulting in more unprocessed ice along the line of sight and thus more CO$_2$:H$_2$O as compared to high-mass YSOs. 

An alternative explanation for the low CO$_2$:H$_2$O fraction toward high-mass stars is different cloud formation time scales and temperature structures during low and high mass star formation. Recent work by \citet{Garrod11} shows that the switchover between CO$_2$:H$_2$O and CO$_2$:CO ice formation depends on temperature and different cloud temperature structures toward low- and high-mass forming clouds may account for the observed differences. The high CO$_2$ abundances toward the small sample of low-mass protostars in a high-radiation region (IC 1396A) seems more consistent with the protostellar processing scenario, however.

In the low-mass protostellar sample there is no correlation between pure CO$_2$ ice abundance and the total envelope mass as traced by H$_2$O ice (\S 3.7.1). There is thus no evidence that variations in pure CO$_2$ fractions can be explained by variable envelope sizes \citep[toward high-mass protostars, envelope temperatures and sizes anti-correlate, see][]{vanderTak00}. Rather large amounts of pure CO$_2$ toward some low-mass protostars may be a remnant of past luminosity outbursts that heated up the envelope beyond its low-accretion temperature, since ice segregation is an irreversible process \citep{Kim11,Kim11b}. Episodic accretion resulting in luminosity outbursts has been invoked to explain the luminosity problem of low-mass YSOs, i.e. the observation that most YSOs are significantly less luminous than predicted by protostellar evolutionary models with steady accretion \citep{Evans09,Dunham10}. 

There is no evidence for a higher CH$_3$OH or OCN$^-$ content toward high-mass stars, except for the cases of overabundant OCN$^-$ toward W33A and very abundant CH$_3$OH toward GL7009S \citep{Dartois99}; thus neither species seems to require stellar UV irradiation or thermal processing to form at typical abundances. The composition of the XCN band varies on average between low- and high-mass objects, but as discussed above the carrier is probably dominated by OCN$^-$ in both cases. Of the remaining ice components C4 and C5 seem to have higher abundances toward the high-mass protostars, indicative of at least a partial formation pathway involving heat or UV irradiation or both, in agreement with the analysis in Paper I. Still, most ice abundances are remarkably similar around low- and high-mass protostars, high-lighting the importance of the cold, protected stages for ice formation up to the complexity of CH$_3$OH. This is also the conclusion in \citet{Boogert11}.

\subsection{Ice evolution in disks?}

\begin{figure}[htp]
\centering
\plotone{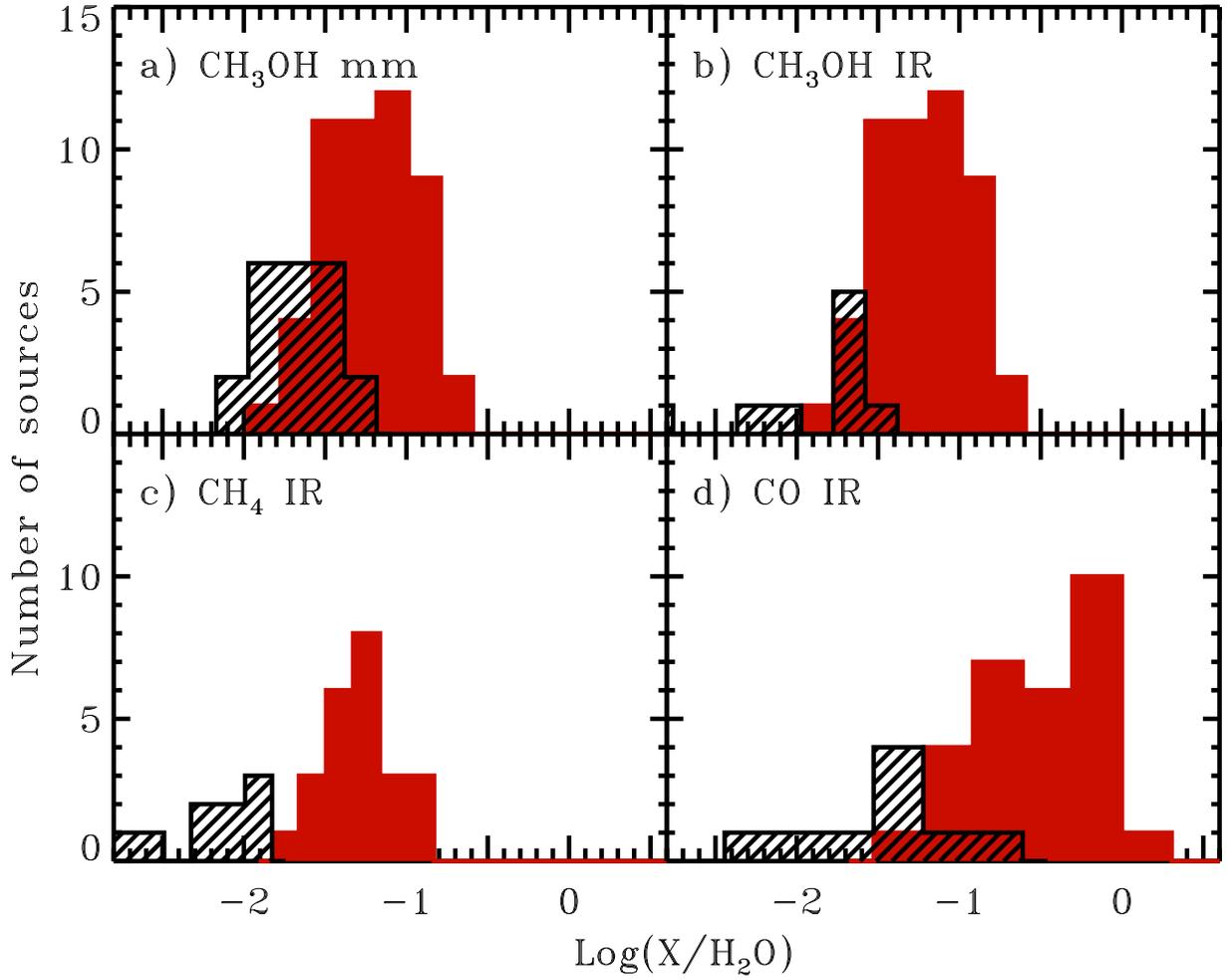}
\caption{Comparison of low-mass protostellar (red solid) and comet (black lines) abundances with respect for water of CH$_3$OH, CH$_4$ and CO. The comet CH$_3$OH abundances in a) are based on millimeter observations from \citet{Biver02}. All other comet abundances are from IR observations compiled by \citet{Disanti08}. \label{fig12}}
\end{figure}

The protostellar stage includes 
the formation of an accretion disk, which is the formation site of planetesimals and eventually planets. These disks are observed to have cold midplanes, where ices should be abundant. It is often difficult to study these disk ices directly because of confusion between disk ices and ices in foreground clouds \citep{Pontoppidan05}. Comets in our own solar system should, however, carry a record of the ice composition in one such disk midplane. It is unknown whether cometary ices reformed in the disk or whether they are at least partly pre-solar or protosolar; recent models suggest that much of the original ice will in fact survive accretion from the envelope into the disk \citep{Visser09}. Even if cometary ices originate from protosolar ices, the protostellar and protosolar ice abundance populations are not expected to overlap perfectly, since the comets originate around a single star, whereas the spread in protostellar ices show the differences in ice abundances between different protostars. Assuming that the protosun resembled one of the observed protostars, what can be addressed is the question of whether comet ice abundances are consistent with a protosolar origin or whether additional ice chemistry in the protosolar nebula is required to account for the observed abundances. In light of this, Figure \ref{fig12} shows a comparison between cometary and low-mass protostellar CH$_3$OH, CH$_4$ and CO abundances  with respect to H$_2$O \citep[][and references therein]{Biver02,Disanti08}. 

For all three ices, the cometary abundances are well below the median protostellar abundances. If the observed cometary ices are pre-solar, this comparison implies either that the ice composition in the protosolar envelope was similar to the most carbon-poor ices observed in the protostellar sample, or that some ice abundances with respect to H$_2$O were reduced in the solar nebula. Possible selective ice destruction processes include desorption of the most volatile ices and UV induced chemistry. 

The spread in cometary ice abundances is comparable to the spread found between {\it different} protostellar sources, suggestive either of a primordial variation in ice abundances within the protosolar envelope or of ice processing in the solar nebula, or a combination of the two. Once a sample of cometary CO$_2$ ice abundances become available, the dominant source of variation should be possible to constrain, since CO$_2$ ice abundances vary little in the protostellar stage. Already, however, the low CH$_4$ abundances do suggest significant ice destruction during the disk stage, through chemistry or selective desorption. Similarly, comets are known to be significantly depleted in nitrogen, and the small number of comets with directly measured NH$_3$/H$_2$O ratios reveal a loss of NH$_3$ compared to low mass protostars \citep[][and references therein]{Kawakita11} -- suggesting that the alteration of cometary ices is not specifically related to carbon. For a more complete cometary ice sample, see upcoming review by Mumma \& Charnley (2011, ARAA review, in press).

%The range of cometary CH$_3$OH abundances are consistent with a prestellar origin, but the range alone suggests that there has been some chemical transformation of the ice in the solar nebula. Whether the chemical reactions have mainly resulted in CH$_3$OH ice formation or destruction is difficult to assess without modeling the ice chemistry in a protoplanetary disk.

\section{Conclusions}

\noindent Large samples of ice sources spanning diverse environments, evolutionary stages and luminosities are necessary both to determine `typical' ice abundances and to identify how ice processes depend on their environment. The main findings from combining the $c2d$ and other ice surveys carried out in a homogenous way are listed below:

\begin{enumerate}
\item  The CO and CO$_2$ abundance medians relative to H$_2$O
are both $\sim$29\% toward low-mass YSOs, while CH$_4$, NH$_3$ and CH$_3$OH are more than a factor of five less abundant. \item CO$_2$:H$_2$O, CH$_4$ and NH$_3$ abundances vary little with respect to H$_2$O, suggesting co-formation. In contrast, CO:H$_2$O, OCN$^-$, CO$_2$:CO and CH$_3$OH vary by factors 2--3 (lower to upper quartile) with respect to H$_2$O, indicative of a separate formation pathway from H$_2$O ice. Pure CO and CO$_2$ ice abundances vary even more, consistent with their sensitivity to protostellar heating.
\item Ice abundances toward background stars and protostars are similar except for a lack of features associated with ice heating, such as pure CO$_2$ ice, toward background stars. In particular, there is no evidence for a different range of CH$_3$OH and CO$_2$ ice abundances. 
\item Protostellar ice abundances in nearby star-forming regions do not vary significantly between different clouds, except for a significantly lower pure CO abundances toward the protostars in Taurus.
\item Compared to low-mass YSOs, ice abundances toward high-mass YSOs are characterized by low levels of CO and CO$_2$ ice, which can be explained by ice heating. Low-mass protostars in the highly irradiated region IC 1396A have typical low-mass protostellar CO$_2$ ice abundances, confirming that the original ice conditions during low- and high-mass star formation are similar. 
\item Correlation studies within the low-mass protostellar sample show a close association between CO, CO$_2$:CO, CO:H$_2$O and the XCN band, supporting the latter's identification with OCN$^-$.
\item Combining the above information, ice formation can generally be divided into three stages: an early phase driven by atomic hydrogenation reactions in clouds, which are fast compared to cloud core formation time scales of $\sim$10$^5$ years;  a later CO-freeze-out dependent ice formation phase which takes place in the pre-stellar phase; and finally the protostellar phase where thermal and possibly UV processing shape the ice content.
\item Toward both low- and high-mass protostars and toward molecular clouds, an average of 8--16\% of the total C, O and N are bound up in ices. Toward the most ice-rich source (IRS 51) 60--80\% of the non-refractory C and O are in ices.
\item There is evidence for more complex ices toward both low- and high-mass protostars, e.g. HCOOH, CH$_3$CHO and/or C$_2$H$_5$OH, but high-resolution spectra toward more sources are required to confirm their presence and quantify their abundances.
\end{enumerate}

{\it Facilities:} \facility{Spitzer, VLT, Keck}

\acknowledgments

Support for K.~I.~O.  is provided by NASA through Hubble Fellowship grant  awarded by the Space Telescope Science Institute, which is operated by the Association of Universities for Research in Astronomy, Inc., for NASA, under contract NAS 5-26555.  Astrochemistry in Leiden is supported by a SPINOZA grant of the Netherlands Organization for Scientific Research (NWO). Support for this work, part of the {\it Spitzer} Space Telescope Legacy Science Program, was provided by NASA through contracts 1224608 and 1230779 issued by the Jet Propulsion Laboratory, California Institute of Technology under NASA contract 1407. The W. M. Keck Observatory is operated as a scientific partnership among the California Institute of Technology, the University of California, and the National Aeronautics and Space Administration, and was made possible by the generous financial support of the W. M. Keck Foundation. The authors  recognize the cultural role and reverence that the summit of Mauna Kea has within the indigenous Hawaiian community. We are most fortunate to have the opportunity to conduct observations from this mountain. We also wish to acknowledge helpful comments from an anonymous referee.

%\bibliographystyle{aa}
%\bibliography{mybib}

\newpage

\newpage

\clearpage
\begin{deluxetable}{lcccccccc}
\tablecolumns{4} 
\tabletypesize{\scriptsize}
\tablecaption{Ice column densities, abundances and optical depths of the 5--7 $\mu$m complex components.
\label{tab2:5-8um}}
\tablewidth{0pt}
\tablehead{
\colhead{Source} & \colhead{N(\water)} & \colhead{[NH$_3$]} & \colhead{[\ch]} &  \colhead{$\tau_{\rm C1}$ (5.84 $\mu$m)}& \colhead{$\tau_{\rm C2}$ (6.18 $\mu$m)}& \colhead{ $\tau_{\rm C3}$ (6.76 $\mu$m)}&  \colhead{$\tau_{\rm C4}$ (6.94 $\mu$m)}& \colhead{$\tau_{\rm C5}$ (broad)}\\
&	10$^{17}$ cm$^{-2}$ &  \% & \%  
}
\startdata
WL 12  & 22.1 $\pm$ 3.0 & $<$3.8 & $<$4.5& 0.012$\pm$0.003 &0.000$\pm$0.002 &0.049$\pm$0.004 &0.136$\pm$0.003 &0.014$\pm$0.071\\
WL 6  & 41.7 $\pm$ 6.0 &  2.9 $\pm$ 0.4 & $<$2.1& 0.002$\pm$0.007 &0.000$\pm$0.006 &0.137$\pm$0.007 &0.087$\pm$0.006 &0.030$\pm$0.045\\
IRS 42  & 19.5 $\pm$ 2.0 &  $<$2.1 & 11.9 $\pm$ 1.1&  -- & -- & -- & -- &--\\
IRS 43 & 31.5 $\pm$ 4.0 & -- & --& 0.059$\pm$0.004 &0.082$\pm$0.003 &0.179$\pm$0.005 &0.105$\pm$0.004 &0.066$\pm$0.051\\
IRS 44  & 34.0 $\pm$ 4.0 & 3.7 $\pm$ 0.4 & $<$1.6& 0.080$\pm$0.005 &0.089$\pm$0.004 &0.135$\pm$0.005 &0.201$\pm$0.004& 0.004$\pm$0.086\\
Elias 32 & 17.9 $\pm$ 2.6 &   $<$5.2  & 12.4 $\pm$ 1.9& 0.021$\pm$0.005 &0.000$\pm$0.004 &0.050$\pm$0.009 &0.075$\pm$0.007 &0.058$\pm$0.025\\
IRS 46  & 12.8 $\pm$ 2.0 &  5.1 $\pm$ 0.9 &  $<$4.1& 0.018$\pm$0.004 &0.002$\pm$0.004 &0.076$\pm$0.005 &0.066$\pm$0.004& 0.000$\pm$0.023\\
VSSG17  & 17.0 $\pm$ 2.5 & $<$3.1 & 6.9 $\pm$ 2.4& 0.058$\pm$0.002 &0.056$\pm$0.002 &0.042$\pm$0.005 &0.061$\pm$0.004 &0.011$\pm$0.016\\
IRS 51  & 22.1 $\pm$ 3.0 &  2.4 $\pm$ 0.3 &  11.7 $\pm$ 0.9&  0.042$\pm$0.003 &0.036$\pm$0.002 &0.074$\pm$0.003 &0.060$\pm$0.002 &0.000$\pm$0.012\\
IRS 63  & 20.4 $\pm$ 3.0 & 5.7 $\pm$ 1.3 &  $<$1.8 & 0.000$\pm$0.003 &0.023$\pm$0.002 &0.048$\pm$0.003 &0.060$\pm$0.003& 0.044$\pm$0.039\\
\enddata
\end{deluxetable}

\clearpage
\begin{deluxetable}{lccc}
\tablecolumns{4} 
\tabletypesize{\scriptsize}
\tablecaption{Typical ice abundances toward low- and high-mass protostars and background stars.  \label{tab:median}}
\tablewidth{0pt}
\tablehead{
\colhead{Ice feature} & \colhead{Low-mass } & \colhead{High-mass }  & \colhead{background}
}
\startdata
      \vspace{0.1cm}
      H$_2$O 		&100 	&100	&100\\
      \vspace{0.1cm}
       CO 			&29		&13		&31\\
      \vspace{0.1cm}
      CO$_2$ 		&29		&13		&38\\
      \vspace{0.1cm}
    CH$_3$OH 		&3  		&4		&4\\
      \vspace{0.1cm}
      NH$_3$ 		& 5 		&5		&\nodata\\
      \vspace{0.1cm}
      CH$_4$ 		& 5		&2		&\nodata\\
      \vspace{0.1cm}
      XCN$^{\rm a}$	& 0.3 	&0.6		&\nodata\\
%       OCN$^-$ 		& 0.2$^{\rm a}$ &0.5$^{\rm a}$	&\nodata\\
\enddata
\\$^{\rm a}$XCN implies a band component securely identified with OCN$^{-}$ and the nearby  2175 cm$^{-1}$ component, which this study suggests is also due to OCN$^{-}$.
\end{deluxetable}

\clearpage
\begin{deluxetable}{lccc}
\tablecolumns{4} 
\tabletypesize{\scriptsize}
\tablecaption{Abundance medians$^{\rm a}$ and lower and upper quartile values of ices and individual ice components with respect to H$_2$O ice \label{tab:median2}}
\tablewidth{0pt}
\tablehead{
\colhead{Ice feature} & \colhead{Low-mass } & \colhead{High-mass }  & \colhead{background}
}
\startdata
      \vspace{0.1cm}
      H$_2$O 		&100 	&100	&100\\
      \vspace{0.1cm}
       CO 			&38$^{61}_{20}$ (29)&13$_{7}^{19}$		&31\\
      \vspace{0.1cm}
      CO$_2$ 		&29$^{35}_{22}$	&13$_{12}^{22}$		&38$_{32}^{41}$\\
      \vspace{0.1cm}
    CH$_3$OH 		&7$^{12}_{5}$ (3)  	&8$_{8}^{16}$ (4)&8$_{7}^{10}$ (4)\\
      \vspace{0.1cm}
      NH$_3$ 		& 5$^{6}_{4}$ 		&16$_{10}^{17}$ (5)	&\nodata\\
      \vspace{0.1cm}
      CH$_4$ 		& 5$^{7}_{4}$		&4$_{2}^{4}$ (2)	&\nodata\\
      \vspace{0.1cm}
      XCN 			& 0.6$^{0.8}_{0.2}$ (0.3) 	&0.8$_{0.4}^{1.4}$ (0.6)	&\nodata\\
      \vspace{0.1cm}
  pure CO 		&21$_{7}^{36}$ 	&3$_{2}^{6}$	&\nodata\\
      \vspace{0.1cm}
   CO:H$_2$O 		&13$_{7}^{19}$ 	&10$_{5}^{12}$	&\nodata\\
      \vspace{0.1cm}
   CO:CO$_2$ 		& 2$_{1}^{3}$ 	&1.3$_{0.4}^{1.6}$ (0.3)	&\nodata\\
      \vspace{0.1cm}
 pure CO$_2$ 		& 2$_{0.3}^{4}$ 	&2$_{1}^{2}$	&\nodata\\
      \vspace{0.1cm}
  CO$_2$:H$_2$O 	&20$_{15}^{23}$	&9$_{6}^{15}$	&24\\
      \vspace{0.1cm}
   CO$_2$:CO 		& 5$_{4}^{7}$	&5$_{2}^{6}$	&6\\
      \vspace{0.1cm}
CO$_2$ shoulder 	& 0.8$_{0.4}^{1.1}$ 	&1$_{1}^{2}$	&\nodata\\
      \vspace{0.1cm}
      OCN$^-$ 		& 0.4$_{0.3}^{0.4}$ (0.2) 	&0.6$_{0.4}^{1.4}$	&\nodata\\
      \vspace{0.1cm}
     2175 cm$^{-1}$	& 0.3$_{0.2}^{0.4}$ (0.2) 	&0.1$_{0.1}^{0.1}$	&\nodata\\
      \vspace{0.1cm}
       C1 (HCOOH + H$_2$CO)$^{\rm b}$ 	& 2.5$_{1.7}^{3.1}$ 	&2.1$_{2.0}^{2.8}$	&2.8$_{2.4}^{3.3}$	\\
      \vspace{0.1cm}
       C2 (HCOO$^{-}$+NH$_3$)$^{\rm b}$ 	& 1.9$_{1.3}^{2.8}$ (1.1)  &1.3$_{1.0}^{1.6}$	&2.5$_{1.3}^{3.3}$\\
      \vspace{0.1cm}
       C3 (NH$_4^+$ + CH$_3$OH)$^{\rm b}$ 	& 4.3 $_{3.0}^{4.7}$	&4.3$_{3.6}^{5.4}$ 	&3.7$_{3.4}^{4.7}$	\\
      \vspace{0.1cm}
       C4 (NH$_4^+$)$^{\rm b}$ 	& 2.3$_{2.1}^{3.7}$  	&4.3$_{2.9}^{5.0}$	&2.1$_{1.1}^{2.8}$\\
      \vspace{0.1cm}
       C5 (warm H$_2$O + anions)$^{\rm b}$ 	&1.5$_{0.8}^{2.2}$ (0.9) 	&3.3$_{0.8}^{6.3}$ (1.4)		&\nodata	\\
\enddata
\\$^{\rm a}$Values in parentheses include upper limits in the median calculation using survival analysis.
\\$^{\rm b}$Since no single carrier the reported number is peak optical depth / ($N_{\rm H_2O}\times 10^{-20} \times 100$).
\end{deluxetable}

\clearpage
\begin{deluxetable}{lccc}
\tablecolumns{4} 
\tabletypesize{\scriptsize}
\tablecaption{Differences in CO and CO$_2$ shoulder medians toward different star forming associations\label{tbl:anova}}
\tablewidth{0pt}
\tablehead{
\colhead{} &\colhead{Pure CO} & \colhead{CO$_2$ shoulder} 
}
\startdata
Ophiuchus	&32		&0.9		\\
Serpens		&38		&1.8		\\
CrA			&29		&1.8		\\
Perseus		&21		&0.8		\\
Taurus		&7		&0.5		\\
Other		&5		&0.5		\\
\tableline
Complete sample	&21		&0.8\\
\enddata
\end{deluxetable}

\clearpage
\begin{deluxetable}{lccccccc}
\tablecolumns{8} 
\tabletypesize{\scriptsize}
\tablecaption{The amount of C, O and N bound up in protostellar ices. \label{tab:con}}
\tablewidth{0pt}
\tablehead{
\colhead{} &\colhead{C$_{\rm ice}$ /  C$_{\rm total}$} &\colhead{O$_{\rm ice}$ /  O$_{\rm total}$} &\colhead{N$_{\rm ice}$ /  N$_{\rm total}$} &\colhead{}&\colhead{C$_{\rm ice}$ /  C$_{\rm vol}$} &\colhead{O$_{\rm ice}$ /  O$_{\rm vol}$}&\colhead{N$_{\rm ice}$ /  N$_{\rm vol}$}   
}
\startdata
Low-mass median	&15\%	&16\%	&10\%	&&27\%	&34\%	&10\%\\
Low-mass max		&46\%	&29\%	&35\%	&&83\%	&61\%	&35\%\\
High-mass median	&8\%	&12\%	&12\%	&&14\%	&25\%	&12\%\\
High-mass max	&18\%	&17\%	&22\%	&&32\%	&36\%	&22\%\\
\enddata
\end{deluxetable}

\newpage

\begin{appendix}

\section{Ophiuchus NH$_3$ and CH$_3$OH spectra}

\begin{figure}
\centering
\epsscale{0.8}
\plotone{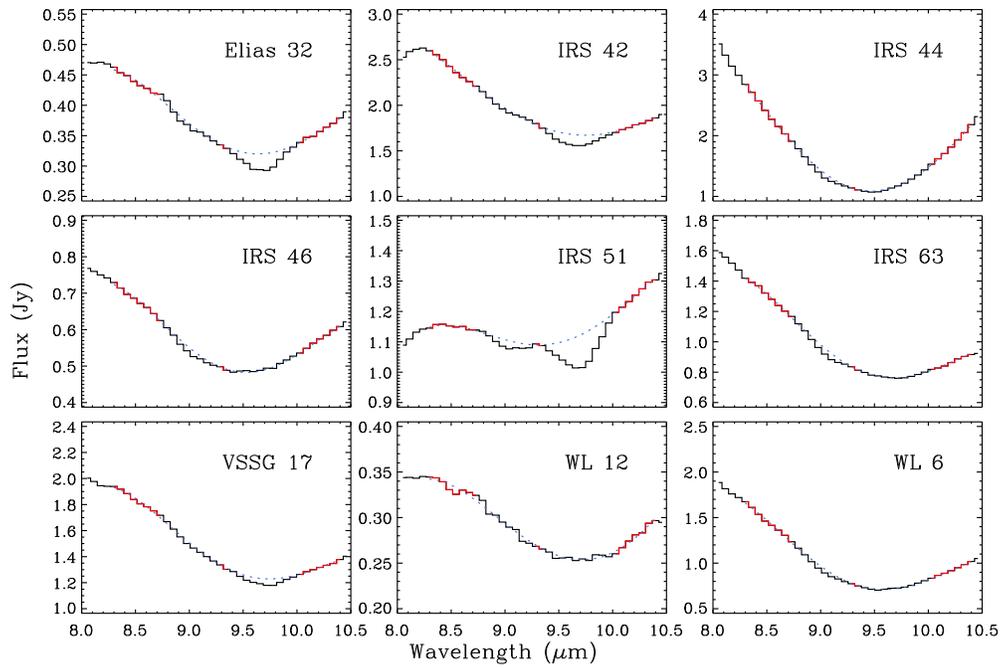}
\caption{{\it Spitzer} spectra between 8.0 and 10.5 $\mu$m showing the silicate feature and the superimposed bands at 9 and 9.7 $\mu$m ascribed to NH$_3$ and CH$_3$OH ice, respectively. The red regions were used to fit a local continuum (blue dotted line). \label{figa1}}
\end{figure}

Figure \ref{figa1} shows the acquired {\it Spitzer} spectra for the nine young stellar objects in Ophiuchus that were added to the $c2d$ sample. The spectra are dominated by a broad silicate feature which is removed by fitting a polynomial to regions free of molecular emission and absorption. We used the same fitting regions as in Paper IV to determine the ice band depths and shapes and then shifted the regions by $\sim$0.1$\mu$m to determine the sensitivity of the results to the chosen local continuum. 

\begin{figure}
\centering
\epsscale{0.8}
\plotone{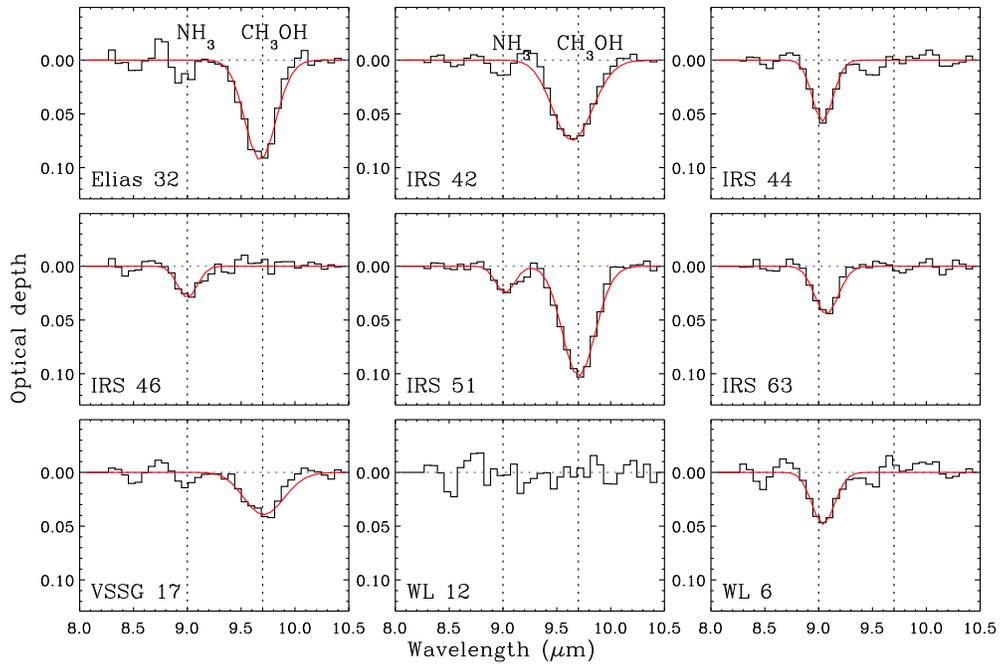}
\caption{Optical depth spectra of NH$_3$ and CH$_3$OH ice features toward the Ophiuchus sources obtained after subtracting the silicate feature. The red line shows the Gaussian fits to the detected bands.\label{figa2}}
\end{figure}

The resulting optical depth spectra are shown in Fig. \ref{figa2}. CH$_3$OH ice is detected toward Elias 32, IRS 42, IRS 51 and VSSG 17 and NH$_3$ ice toward IRS 44, IRS 46, IRS 51, IRS 63 and WL 6. The bands are fitted with Gaussians and the derived peak positions and full-width-half-maxima are listed in Table \ref{tab:a1}. The uncertainties are dominated by the choice of local continuum. The NH$_3$ peak positions are consistent with laboratory ice spectra, while the FWHM are generally too low. From Paper IV, we know that the NH$_3$ FWHM may be underestimated when using a local continuum rather than a silicate template to extract optical depth spectra and this explains the discrepancy. The CH$_3$OH peak positions and FWHM are both consistent with laboratory measurements, with the peak positions suggesting that CH$_3$OH is present in a CO-rich ice (Paper IV).

\begin{table}[h]
{\footnotesize
\caption{\fw\ and peak positions of detected \nhh\ and \ch\ bands compared to laboratory values.}
\label{tab:a1}
\begin{center}
\begin{tabular}{ l c c c c c c  }
\hline
\hline
Source & \multicolumn{2}{c}{Peak position \nhh\ } & \fw\ \nhh\ & \multicolumn{2}{c}{Peak position \ch\ }& \fw\ \ch\  \\
&$\mu$m &	cm$^{-1}$& cm$^{-1}$ & $\mu$m & cm$^{-1}$  & cm$^{-1}$ \\
\hline
Elias 32 & \nodata & \nodata & \nodata & 9.68$\pm0.01$&1033$\pm$1&36 $\pm$ 3 \\
IRS 42 &  \nodata & \nodata & \nodata &9.64$\pm$0.01 &1037$\pm$1 &47 $\pm$ 3 \\
IRS 44 & 9.04$\pm$0.01&1107$\pm$1& 28 $\pm$ 2 &  \nodata & \nodata & \nodata \\
IRS 46 & 9.00$\pm$0.01 &1110$\pm$1& 27 $\pm$ 3 &\nodata & \nodata & \nodata \\
IRS 51 & 9.02$\pm$0.01 &1109$\pm$1 & 25 $\pm$ 1 & 9.70$\pm$0.01 &1031$\pm$1& 38 $\pm$ 2 \\
IRS 63 & 9.07$\pm$0.02 &1102$\pm$2& 32 $\pm$ 5 & \nodata & \nodata & \nodata \\
VSSG 17 & \nodata & \nodata & \nodata&9.71$\pm$0.02&1030$\pm$3 & 45 $\pm$ 10 \\
WL 12 & \nodata & \nodata & \nodata&\nodata & \nodata & \nodata \\
WL 6 & 9.04$\pm$0.01 &1106$\pm$1 &31 $\pm$ 2 & \nodata & \nodata & \nodata \\ 
\hline
Laboratory$^{\rm a}$ &8.85--9.41&1062--1130&53--137&9.67--9.83&1017--1034&22--39\\
\hline
\end{tabular}
\end{center}
$^{\rm a}$\citet{Bottinelli10}
}
\end{table}

\newpage

\section{Carriers of the 7.25 $\mu$m ice feature}

Paper I showed that it is difficult to fit the 7.25 $\mu$m with spectra of HCOOH, its most commonly assigned carrier. Laboratory spectra of some tertiary mixtures come closest, but suffer from uncertainties in baseline determinations (see below).

Figure \ref{figa3} shows the spectra of pure HCOOH ice and an H$_2$O:HCOOH 4:1 ice mixture \citep[at 15~K from][]{Bisschop07b}), pure CH$_3$CH$_2$OH and CH$_3$CHO spectra \citep{Oberg09d}, and observed spectra toward W33A, NGC7538 IRS9 and B1-b. W33A and B1-b clearly have absorption features at 7.25 and 7.4~$\mu$m, while NGC7538 IRS9 does not. The comparison shows that CH$_3$CH$_2$OH is a plausible carrier for the 7.25 $\mu$m feature, commonly assigned to HCOOH. The lack of the 7.25~$\mu$m toward  NGC7538 IRS9 is consistent with the upper limit on CH$_3$CH$_2$OH ice from the 3~$\mu$m feature \citep{Boudin98}, while no such limits are published for the other sources. Figure \ref{figa3} also shows that the observed 7.40 $\mu$m feature is likely due to CH$_3$CHO. Thus there is also little evidence for HCOO$^-$ -- both have been proposed previously as carriers \citep[e.g.][Paper I]{Schutte99,Gibb04}. 

These tentative identifications may in the future provide an opportunity to observe a complex ice chemistry {\it in situ}, but the Spitzer-IRS spectra are too low in spectral resolution to securely distinguish between different carriers, especially since the HCOOH spectral feature depends on the ice matrix and has been reported to become as narrow as the 
CH$_3$CH$_2$OH band in mixtures with H$_2$O and CH$_3$OH, but this result depends crucially on the choice of local baseline \citep{Bisschop07b}. Until the HCOOH baseline issue has been resolved and higher resolution spectra exist for more sources the HCOOH and HCOO$^-$ abundances should not be determined from the 7.25 and 7.42 $\mu$m ice features. This does not imply that HCOOH and HCOO$^-$ are not present in interstellar ices, only that we do not currently have the tools to quantify their abundances.

\begin{figure}
\centering
\epsscale{0.5}
\plotone{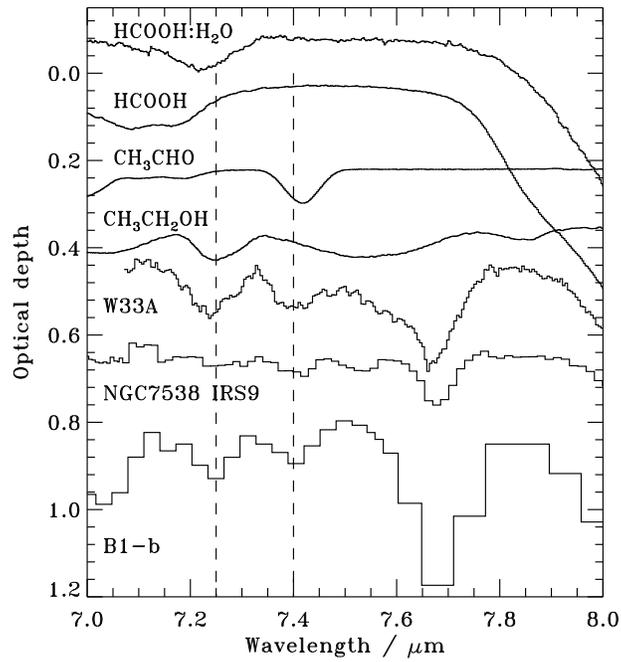}
\caption{\textit{ISO} spectra at 7--8 $\mu$m for W33A, NGC7538 IRS9 and B1-b, following subtraction of a local spline continuum, plotted together with laboratory spectra of pure HCOOH, CH$_3$CHO and CH$_3$CH$_2$OH ices. The dashed lines mark the 7.25 and 7.40 $\mu$m features usually assigned to HCOOH and HCOO$^-$. The feature at 7.67 $\mu$m is due to CH$_4$ ice.  The baselines were obtained by fitting a local spline continuum to 6.9, 7.16, 7.33, 7.77 and 7.85 $\mu$m.\label{figa3}}
\end{figure}

\newpage

\section{Histograms of ice detections and detections$+$upper limits}

\begin{figure}[htp]
\centering
\plotone{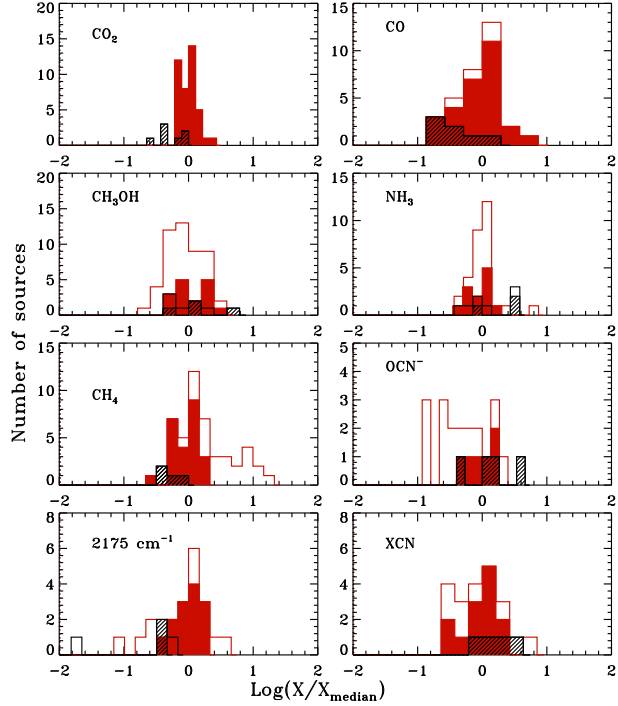}
\caption{Histograms of total ice abundances toward low-mass protostars (red) and high-mass protostars (black). Solid histograms mark detections and contours detections$+$upper limits. For low-mass NH$_3$ the solids are from Paper IV using a silicate template to extract NH$_3$ and the contour are from Paper IV and Table 1 when using a polynomial to fit the local continuum. \label{fig16}}
\end{figure}

\begin{figure}[htp]
\centering
\plotone{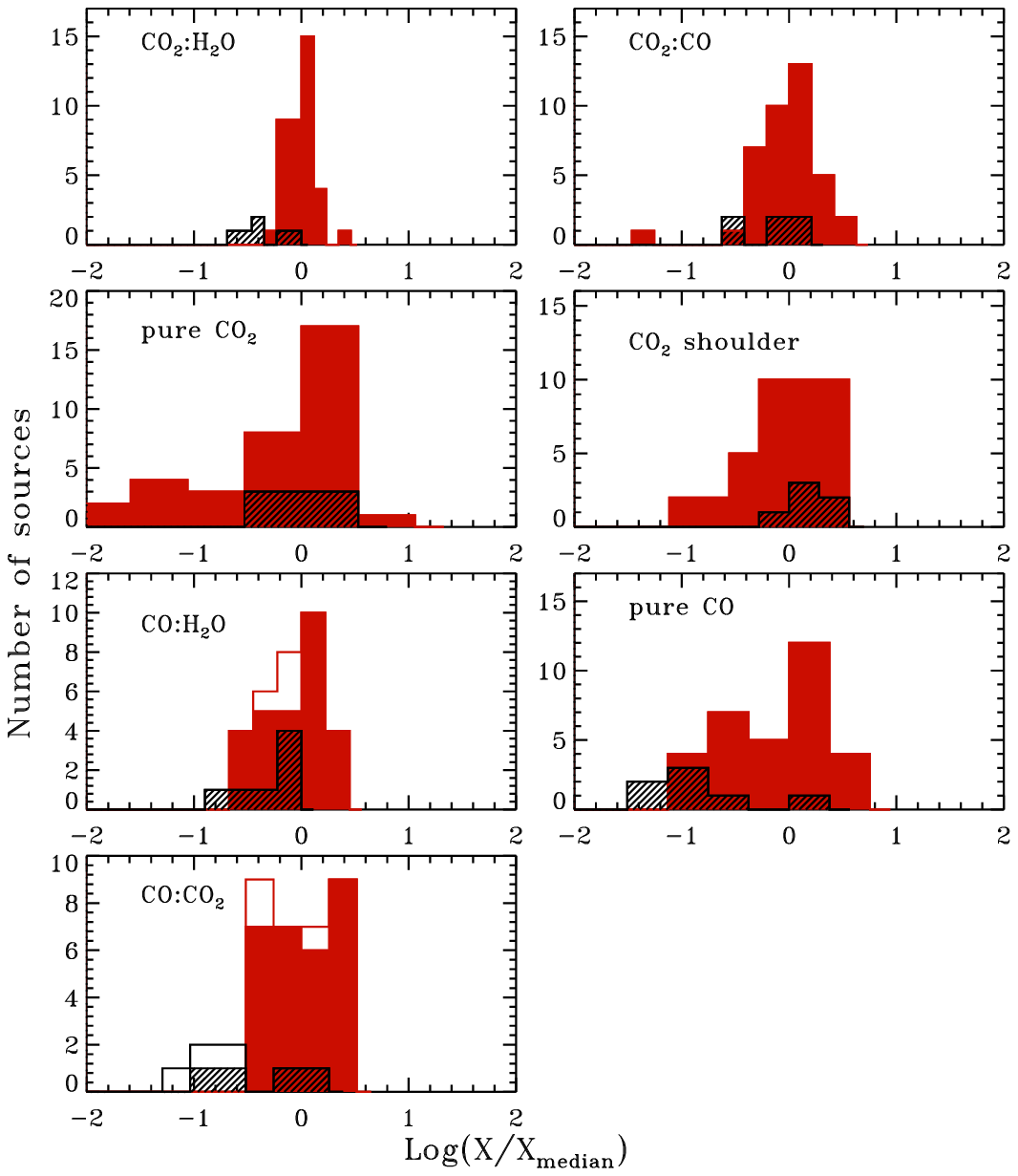}
\caption{Histograms of CO and CO$_2$ components, otherwise as in Fig. \ref{fig16} \label{fig17}}
\end{figure}

\begin{figure}[htp]
\centering
\plotone{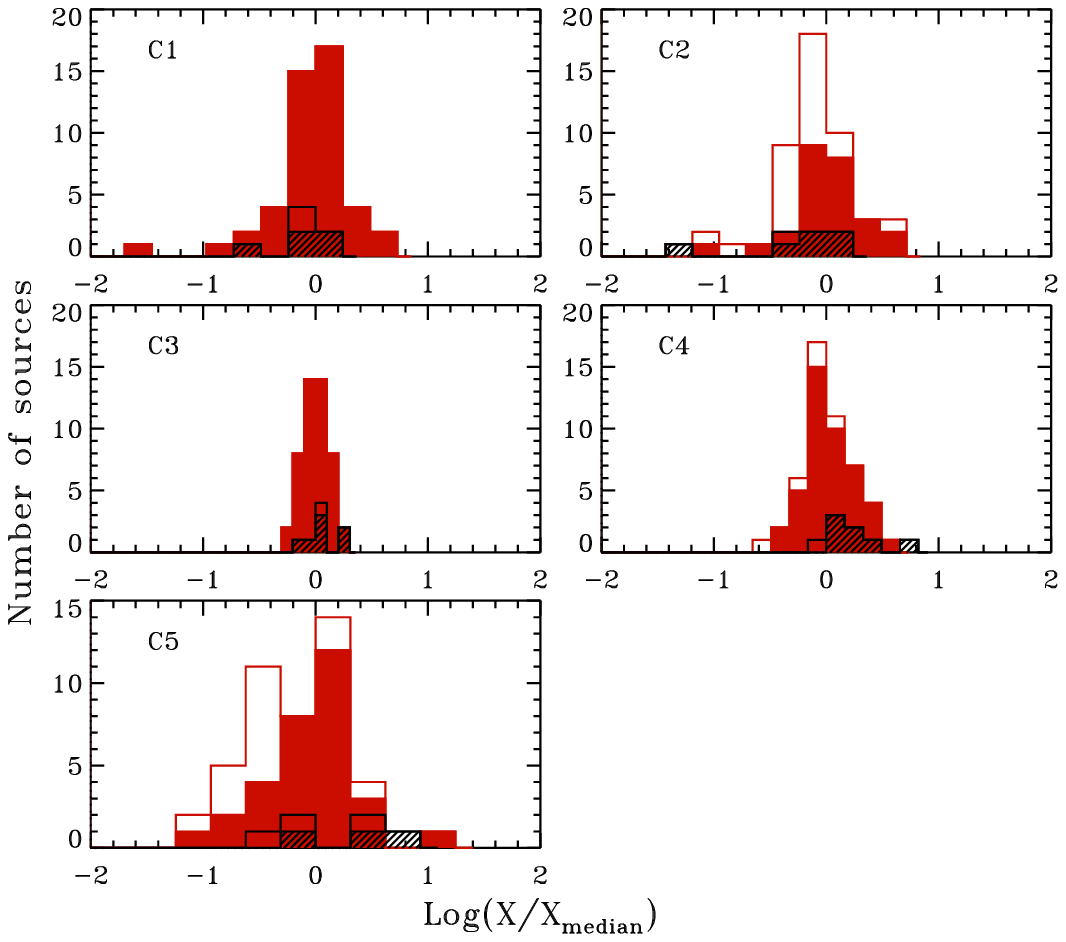}
\caption{Histograms of 5-7 $\mu$m components, otherwise as in Fig. \ref{fig16}. \label{fig18}}
\end{figure}

\end{appendix}

\end{document}